\documentclass[twocolumn]{aastex631}
\usepackage{amsmath}

\usepackage{color}

\newcommand{\hi}{H\textsc{i}}
\providecommand{\sorthelp}[1]{}

\begin{document}

\title{The Astrodust+PAH Model:\\ A Unified Description of the Extinction, Emission, and Polarization from Dust in the Diffuse Interstellar Medium}

\author[0000-0001-7449-4638]{Brandon S. Hensley}
\email{bhensley@astro.princeton.edu}
\affiliation{Department of Astrophysical Sciences,  Princeton University, Princeton, NJ 08544, USA}
\affiliation{Spitzer Fellow}

\author[0000-0002-0846-936X]{B. T. Draine}
\affiliation{Department of Astrophysical Sciences,  Princeton University, Princeton, NJ 08544, USA}

\date{\today}

\begin{abstract}
We present a new model of interstellar dust in which large grains are a single composite material, ``astrodust,'' and nanoparticle-sized grains come in distinct varieties including polycyclic aromatic hydrocarbons (PAHs). We argue that a single-composition model for grains larger than $\sim0.02\,\mu$m most naturally explains the lack of frequency dependence in the far-infrared (FIR) polarization fraction and the characteristic ratio of optical to FIR polarization. We derive a size distribution and alignment function for 1.4:1 oblate astrodust grains that, with PAHs, reproduce the mean wavelength dependence and polarization of Galactic extinction and emission from the diffuse interstellar medium while respecting constraints on solid phase abundances. All model data and Python-based interfaces are made publicly available.
\end{abstract}

\section{Introduction}
What is the nature of interstellar dust? The elemental building blocks of dust grains are, for the most part, well-established. Comparing gas phase depletions \citep{Jenkins_2009} to Solar abundances \citep{Asplund_2021, Magg+etal_2022} and accounting for diffusion corrections \citep{Turcotte+Wimmer-Schweingruber_2002} and the enrichment of interstellar gas with metals since the formation of the Sun \citep{Chiappini+Romano+Matteucci_2003, Bedell+etal_2018}, one can infer the amount of each element depleted onto dust. We have recently derived solid phase abundances per H atom of $126\pm56$\,ppm for C, $249\pm94$\,ppm for O, $45.8\pm4.9$\,ppm for Mg, $3.4\pm0.3$\,ppm for Al, $38.0\pm3.1$\,ppm for Si, $7.6\pm2.0$\,ppm for S, $3.2\pm0.2$\,ppm for Ca, $42.8\pm4.0$\,ppm for Fe, and $2.0\pm0.2$\,pm for Ni on diffuse, high-latitude sightlines \citep{Hensley+Draine_2021}. Other determinations from similar underlying data are in broad agreement \citep[e.g.,][]{Jenkins_2009, Compiegne+etal_2011, Draine_2011, Zuo_2021}.

The chemical composition of dust is more uncertain, but is constrained by spectroscopic features in extinction and emission \citep[see][for a recent overview]{Hensley+Draine_2021}. For instance, the 2175\,\AA\ feature can be explained by a plausible abundance of $sp^2$-bonded carbon \citep{Stecher+Donn_1965} while a series of mid-infrared features suggest a large population of polycyclic aromatic hydrocarbons \citep[PAHs;][]{Leger+Puget_1984, Allamandola+Tielens+Barker_1985}, perhaps the same grains as the carriers of the 2175\,\AA\ feature \citep{Joblin+Leger+Martin_1992,Steglich_2011}. Aliphatic hydrocarbons are well-attested by mid-infrared features at 3.4 and 6.85\,$\mu$m \citep{Tielens+etal_1996,Chiar+etal_2000,Hensley+Draine_2020}. The 9.7 and 18\,$\mu$m features have long been identified with amorphous silicates \citep{Woolf+Ney_1969,vanBreemen+etal_2011}, and minimal fine structure in these features in the diffuse interstellar medium (ISM) limits the crystalline fraction to $\lesssim2\%$ \citep{DoDuy+etal_2020}. It has been argued that the profile shapes are more consistent with Mg-rich silicates than Fe-rich ones \citep[e.g.,][]{Min+etal_2007, vanBreemen+etal_2011, Poteet+Whittet+Draine_2015}, leaving indeterminate in what form the bulk of the solid phase Fe resides.

Even having established that interstellar dust consists of, at minimum, aromatic and aliphatic hydrocarbons and amorphous silicates, a central question remains: what is the {\it macroscopic} composition of grains? Is a typical grain predominantly either hydrocarbon or silicate in nature, or is it a composite material made up of both materials (and likely others)? The goal of this work is to put forward a physical model of interstellar dust in which grains are assumed to be a composite material, ``astrodust'' \citep{Draine+Hensley_2021a}, in contrast with most dust models developed over the past four decades that have invoked distinct silicate and carbonaceous populations \citep[e.g.,][]{Mathis+Rumpl+Nordsieck_1977,Draine+Lee_1984,Weingartner+Draine_2001,Zubko+Dwek+Arendt_2004,Compiegne+etal_2011,Jones+etal_2017,Siebenmorgen_2017,Guillet+etal_2018}.

How the ISM could maintain separate populations of silicate and carbonaceous grains is not clear. The dust residence time is long compared to grain destruction timescales, suggesting that a typical grain will have been broken down in shocked gas and regrown via gas-phase accretion following its injection into the ISM as stardust. While such arguments have been previously advanced \citep[e.g.,][]{Draine+Salpeter_1979, Dwek+Scalo_1980, Seab+Shull_1983}, some authors \citep[e.g.,][]{Jones+Nuth_2011} have argued for lower grain destruction rates, so that stardust grains can account for a substantial fraction of the interstellar grain mass. Even if stardust grains are substantially destroyed, it is possible that selective regrowth in the ISM might maintain separate silicate and carbonaceous grains \citep{Draine_1990, Draine_2009}. The fact that such two-component models proved quite successful at reproducing observations of the wavelength-dependence of interstellar extinction and emission led to their widespread use in modeling interstellar dust. While a few earlier studies developed models in which large grains are a single population of core-mantle particles and small grains have diverse compositions \citep{Hong+Greenberg_1980, Li+Greenberg_1997}, one-component models have otherwise been relatively unexplored (though see discussions by \citet{Kim_1994} and \citet{Mathis_1996}).

Recently, observational evidence favoring a one-component model of interstellar dust has come to light in the form of far-infrared (FIR) polarimetry. The optical properties and polarization efficiencies of silicate and carbonaceous grains are likely different, and consequently the dust spectral energy distribution (SED) in total intensity is expected to differ from that in polarized intensity \citep[see discussion in][]{Draine+Hensley_2021a}. Indeed, \citet{Draine+Fraisse_2009} predicted that the dust polarization fraction should increase with increasing wavelength in the FIR as the more polarizing silicates contributed an increasing fraction of the total emission. However, observations from the Planck satellite \citep{Planck_2018_XI} and those from the BLASTPol balloon \citep{Ashton+etal_2018, Shariff+etal_2019} constrain the frequency dependence of the polarization fraction on diffuse sightlines to deviations of $\lesssim10$\% from a constant value from 250\,$\mu$m to 3\,mm. While two-component dust models can be constructed to comply with these constraints if the FIR optical properties of the components are sufficiently similar \citep{Guillet+etal_2018}, the need for fine-tuning prompts a questioning of the two-component paradigm.

In this paper, we construct a unified model of the total and polarized extinction and emission of dust in the diffuse ISM based on astrodust and PAHs. In Section~\ref{sec:modeling}, we discuss calculation of the extinction, emission, and polarization from an ensemble of partially aligned spheroidal grains; in Section~\ref{sec:materials}, we describe the grain materials employed; in Section~\ref{sec:data_model}, we detail the framework used to derive the size distributions and alignment function of astrodust grains and PAHs that best reproduces the observational data; in Section~\ref{sec:results}, we present the best-fit model and evaluate the goodness of fit; in Section~\ref{sec:discussion}, we discuss the implications of our results on the nature of interstellar grains and comment on future theoretical and observational directions; finally, in Section~\ref{sec:summary} we summarize our principal conclusions.

\section{Extinction, Emission, and Polarization from Partially Aligned Grains} \label{sec:modeling}

\subsection{Cross Sections}
\label{subsec:cross_sections}

The observed polarized extinction and emission from interstellar dust implies that the grains are aspherical and aligned, requiring models of dust polarization to go beyond Mie theory calculations with spherical particles. Here we adopt the next simplest approximation: spheroidal grains.

Let $a_1$, $a_2$, and $a_3$ denote the principal semi-axes of a spheroidal grain, where $a_1$ is the symmetry axis. We focus principally on oblate spheroids, which have $a_1 < a_2 = a_3$. We define the ``effective radius'' $a$ of the spheroid as the radius of a sphere having equal volume, i.e.,

\begin{equation}
    a \equiv (a_1 a_2 a_3)^{1/3}
    ~~~.
\end{equation}

Let $\hat{\bf a}_1$ be a unit vector along $a_1$, $\hat{\bf k}$ be the direction of propagating photons, and $\Theta$ be the angle between $\hat{\bf k}$ and $\hat{\bf a}_1$. The absorption, scattering, and extinction cross sections then depend both on $\Theta$ and the relative orientation of the photon electric field ${\bf E}$ and $\hat{\bf a}_1$. Following \citet{Draine+Hensley_2021c}, we define $C_E(\Theta)$ as the cross section (any of absorption, scattering, or extinction) corresponding to photons with ${\bf E}$ in the $\hat{\bf k}$--$\hat{\bf a}_1$ plane and $C_H(\Theta)$ as the cross section corresponding to the photon magnetic field ${\bf H}$ in the $\hat{\bf k}$--$\hat{\bf a}_1$ plane.

To compute the absorption, scattering, or emission from a population of arbitrarily oriented grains, we would in principle need to compute $C_E$ and $C_H$ for all values of $\Theta$ and then average over grain rotation and precession. In practice, the orientational averaging can be approximated by considering only the three grain orientations corresponding to the grain symmetry axis oriented along one of three orthogonal directions. This ``modified picket fence approximation'' \citep[MPFA;][]{Draine+Hensley_2021c} is effectively a 3D extension of the original ``picket fence'' approximation \citep{Dyck+Beichman_1974}. While the MPFA becomes exact in the limit of wavelengths $\lambda \gg a$, even at wavelengths comparable to the grain size where the errors are largest ($\sim$ 10\%), the MPFA is suitably accurate for calculations of dust polarization when averaging over a realistic grain size distribution \citep{Draine+Hensley_2021c}.

The three MPFA orientations correspond to $C_E(0)$, $C_E(90^\circ)$, and $C_H(90^\circ)$, where $C_E(0) = C_H(0)$ from the spheroidal symmetry. With these definitions, the MPFA cross section $C_{\rm ran}$ for randomly oriented grains is

\begin{equation} \label{eq:cran}
    C_{\rm ran} = \frac{1}{3}\left[C_E(0) + C_E(90^\circ) + C_H(90^\circ)\right]
    ~~~,
\end{equation}
where $C$ can refer to absorption, scattering, or extinction. The MPFA polarization cross section for oblate spheroids is 

\begin{equation} \label{eq:cpol}
    C_{\rm pol}^{\rm oblate} = \frac{1}{2}\left[C_H(90^\circ) - C_E(90^\circ)\right]
    ~~~.
\end{equation}

The quantities $C_{\rm ran}$ and $C_{\rm pol}$ as a function of grain size and wavelength are sufficient to characterize the cross sections for each grain component in our model and to compute the observables of interest. We employ a precomputed library of MPFA cross sections for spheroids with sizes uniformly spaced in $\log(a)$ (in steps $\Delta\log_{10}(a)=0.025$) and wavelengths uniformly spaced in $\log(\lambda)$ (in steps $\Delta\log_{10}(\lambda)=0.005$).

\subsection{Extinction}

Let each grain component $i$ in our model have a size distribution $dn_i/da$ defined such that the number density of grains having an effective radius between $a$ and $a+da$ is $\left(dn_i/da\right)da$. Then the total extinction $\tau_\lambda$ per H column density $N_{\rm H}$ at wavelength $\lambda$ for a population of randomly oriented grains is

\begin{equation} \label{eq:ext}
\frac{\tau_\lambda}{N_{\rm H}} = \sum\limits_{i = 1}^N
\int da \left(\frac{1}{n_{\rm H}}\frac{dn_i}{da}\right) C^{\rm ext}_{\rm ran}\left(\lambda, i,
  a\right)
~~~,
\end{equation}
where $N$ is the number of distinct components in the model and $C^{\rm ext}_{\rm ran}$ denotes the extinction cross section for randomly oriented grains (Equation~\eqref{eq:cran}). We compute $\tau_\lambda/N_{\rm H}$ for a population of randomly oriented rather than partially aligned grains as we compare against observations averaged over many sightlines and thus grain orientations. 

The polarized extinction per $N_{\rm H}$ depends upon the extinction cross sections, the degree of alignment of the grains, and the relative orientation of the grain alignment direction and the line of sight. We assume that grains tend to rotate about their axis of greatest moment of inertia, i.e., their short axis, and that the rotation axis is preferentially aligned with the local magnetic field. However, alignment between the grain angular momentum, the short axis, and magnetic field is imperfect, with grains undergoing both free and magnetic precession. The alignment efficiency $f_{\rm align}$ quantifies the time-averaged degree of alignment between the grain short axis $\hat{\bf a}_1$ and the magnetic field as a function of grain size:

\begin{equation}
    f_{\rm align} \equiv \frac{3}{2} \left[\Bigl\langle\left(\hat{\bf a}_1 \cdot \hat{\bf B}\right)^2 \Bigr\rangle - \frac{1}{3}\right]
    ~~~,
\end{equation}
where $\hat{\bf B}$ is a unit vector along the local magnetic field direction. $f_{\rm align} = 0$ corresponds to randomly oriented grains and $f_{\rm align} = 1$ to perfect spinning alignment. It is analogous to the reduction factor defined by \citet{Greenberg_1968}.

We aim in this work to reproduce the maximum dust polarization per $N_{\rm H}$, corresponding to the magnetic field lying in the plane of the sky. In this limit and using the MPFA, the polarized extinction per $N_{\rm H}$ from a population of partially aligned grains is

\begin{equation}
    \left(\frac{p_\lambda}{N_{\rm H}}\right)^{\rm max} = \sum\limits_{i = 1}^N
\int da \left(\frac{1}{n_{\rm H}}\frac{dn_i}{da}\right) C^{\rm ext}_{\rm pol}\left(\lambda, i,
  a\right) f_{\rm align}\left(i,a\right)
  ~~~,
\end{equation}
where $C^{\rm ext}_{\rm pol}$ denotes the polarization cross section for extinction (Equation~\eqref{eq:cpol}).

\subsection{Emission}
There are two principal emission mechanisms for interstellar grains: thermal vibrational emission $I_\lambda^{\rm th}$ and rotational emission $I_\lambda^{\rm SpD}$ (``spinning dust emission''). The total dust emission per H atom $I_\lambda/N_{\rm H}$ is the sum of these two contributions, i.e.,

\begin{equation}
\frac{I_\lambda}{N_{\rm H}} = \frac{I_\lambda^{\rm th}}{N_{\rm H}} + \frac{I_\lambda^{\rm SpD}}{N_{\rm H}}
~~~.
\end{equation}
We discuss each in turn below. In this work we do not attempt to model the dust luminescence that has been observed at optical wavelengths \citep[see][for a recent discussion]{Witt+Lai_2020}.

\subsubsection{Thermal Vibrational Emission} \label{subsubsec:theory_irem}
Interstellar grains absorb UV and optical starlight and reradiate the energy in the infrared. The thermal vibrational emission per $N_{\rm H}$ from a population of randomly oriented dust grains is

\begin{align}
\frac{I_\lambda^{\rm th}}{N_{\rm H}} = \sum\limits_{i = 1}^N \int da &\left(\frac{1}{n_{\rm H}}\frac{dn_i}{da}\right)\times \nonumber \\
&\int dT \left(\frac{dP}{dT}\right)_{i,a} C^{\rm abs}_{\rm ran}\left(\lambda, i, a\right) B_\lambda\left(T\right)
~~~,
\end{align}
where $B_\lambda\left(T\right)$ is the Planck function and $(dP/dT)_{i,a}$ is the grain temperature distribution of grains of composition $i$ and effective radius $a$. We ignore the contribution from stimulated emission, which is negligible in radiation fields typical of the diffuse ISM \citep{Li+Draine_2001b}.

To compute the temperature distribution of a grain of composition $i$ and effective radius $a$, we employ the methods of \citet{Guhathakurta+Draine_1989} and \citet{Li+Draine_2001} assuming a radiation field with specific energy density $u_\lambda$. We base the functional form of $u_\lambda$ on the determination of the local interstellar radiation field by \citet{Mathis+Mezger+Panagia_1983} (MMP) with updates following \citet{Draine_2011}, i.e., the ``modified MMP'' (mMMP) starlight spectrum. Specifically, we take

\begin{equation} \label{eq:U}
u_\lambda = U\left[u_\lambda^{\rm UV} + \sum_{i=1}^3 \frac{4\pi}{c}
  W_i B_\lambda\left(T_i\right)\right] +
\frac{4\pi}{c}B_\lambda\left(T_{\rm CMB}\right)
~~~,
\end{equation}
where $U$ is a frequency independent scaling factor, $u_\lambda^{\rm UV}$ is the UV component of the radiation field, and $T_{\rm CMB} = 2.725\,$K is the CMB temperature \citep{Fixsen_2009}. The optical starlight component is modeled as three blackbodies having temperatures of $T_1 = 3000$\,K, $T_2 = 4000$\,K, and $T_3 = 7500$\,K with dilution factors $W_1 = 7\times10^{-13}$, $W_2 = 1.65\times10^{-13}$, and $W_3 = 1\times10^{-14}$. The UV component is given by

\begin{align}
&\frac{\lambda u_\lambda^{\rm UV}}{\rm erg\,cm^{-3}} = \nonumber \\
&\begin{cases}
2.373\times10^{-14}\left(\frac{\lambda}{\mu{\rm m}}\right)^{-0.6678}
& 1340\, \text{\AA}\ < \lambda < 2460\,\text{\AA} \\
6.825\times10^{-13}\frac{\lambda}{\mu{\rm m}}
& 1100\, \text{\AA}\ < \lambda < 1340\, \text{\AA} \\
1.287\times10^{-9}\left(\frac{\lambda}{\mu{\rm m}}\right)^{4.4172}
& 912\, \text{\AA}\ < \lambda < 1100\, \text{\AA} \\
0 & \lambda < 912\, \text{\AA}
~~~.
\end{cases}
\end{align}
The IR emission from the dust itself does not appreciably contribute to the heating of dust grains, and we neglect it here.

To ensure consistency between the power absorbed by dust having the adopted extinction curve and the power emitted by dust having the observed SED, we require $\log_{10} U = 0.2$ ($U \simeq 1.6$). A factor $\langle U \rangle \sim 1.2$ was derived by \citet{Draine+Li_2007} based on the FIR SED derived by \citet{Finkbeiner+Davis+Schlegel_1999}, where the FIR emission was dominated by dust heated by a $U = 0.8$ spectrum. While several factors contribute to the larger value of $U$ adopted in the present study relative to \citet{Draine+Li_2007}, including refinement of constraints on FIR emission and differences in the dust model albedos at optical and UV wavelengths, the most significant are the systematically larger submillimeter opacities of astrodust (see Figure~\ref{fig:opacities}) and our adoption of a value of $A_V/N_{\rm H}$ that is about one third lower (see Section~\ref{subsec:model_compare}). Since the shape of the dust SED constrains the grain temperature to $\sim 18$\,K, our larger submillimeter opacities necessitated a more intense heating spectrum to maintain comparable grain temperatures. Likewise, our adoption of a lower value of $E(B-V)/N_{\rm H}$ without a correspondingly smaller $I_\lambda/N_{\rm H}$ could only be accommodated by an increase in $U$.

Just as a population of aligned, aspherical grains produces polarized extinction, the emission from these grains is polarized. For partially aligned spheroidal grains assuming the MPFA, the maximum polarized thermal emissivity per H atom is given by

\begin{align}
\label{eq:ipol}
\left(\frac{P_\lambda}{N_{\rm H}}\right)^{\rm max} &= \sum\limits_{i = 1}^N \int da \left(\frac{1}{n_{\rm H}}\right) \frac{dn_i}{da} \int dT \left(\frac{dP}{dT}\right)_{i,a}\times \nonumber \\ 
&f_{\rm align}\left(i, a\right) C_{\rm pol}^{\rm abs}\left(\lambda, i, a\right) B_\lambda\left(T\right)
~~~,
\end{align}
corresponding to a viewing geometry with the magnetic field in the plane of the sky.

\subsubsection{Spinning Dust Emission}
In addition to vibrational emission, rapidly rotating ultrasmall grains can also produce electric or magnetic dipole emission. This mechanism has been identified with the AME peaking near 30\,GHz \citep[see][for a review]{Dickinson+etal_2018}. To compute the spinning dust emissivity $j_\lambda^{\rm AME}$, we employ the \texttt{SpDust} software \citep{Ali-Haimoud+Hirata+Dickinson_2009, Silsbee+AliHaimoud+Hirata_2011} with extension to silicate grains by \citet{Hensley+Draine_2017b} and with dust properties as described in Section~\ref{sec:materials}. We amend the \texttt{SpDust} default mass density parameters for both PAHs and silicates to ensure consistency with those adopted in the present study for PAHs (2.0\,g\,cm$^{-3}$) and astrodust (2.74\,g\,cm$^{-3}$), respectively.

We adopt the environmental parameters of \citet{Draine+Lazarian_1998a} for computing spinning dust emission from the warm neutral medium (WNM) and cold neutral medium (CNM). However, we increase the radiation field parameter $\chi$ from 1 to 1.6 to better agree with the spectrum of the interstellar radiation field described in Section~\ref{subsubsec:theory_irem}. Following observations of \hi\ absorption in the diffuse ISM by the 21-SPONGE survey \citep{Murray+etal_2020}, we adopt a fiducial CNM fraction of 28\% to compute the weighted average between the WNM and CNM spinning dust spectra. There is evidence that the most diffuse lines of sight at high Galactic latitudes can have even lower CNM fractions \citep{Heiles+Troland_2003, Murray+etal_2020, Hensley+Murray+Dodici_2022}, and we find that with an appropriate distribution of electric dipole moments we can obtain satisfactory fits assuming lower values for the CNM fraction, e.g., 5\%.

Rotational electric or magnetic dipole radiation from a single spinning grain is completely polarized, and so the AME is expected to be polarized if the ultrasmall grains responsible for it are systematically aligned. \citet{Lazarian+Draine_2000} argued that ``resonant paramagnetic relaxation'' could lead to modest degrees of alignment for nanoparticles, with $\sim2$\% fractional polarization at 20\,GHz \cite[see also][]{Hoang+Lazarian+Martin_2014}. However, more recently we have argued that all alignment processes are suppressed for grains small enough to produce rotational emission at tens of GHz on account of quantization of their rotational and vibrational energy levels \citep{Draine+Hensley_2016}. Empirically, no AME polarization has been definitively observed, with upper limits of order a few percent \citep{Macellari+etal_2011, Planck_2015_XXV, GenovaSantos+etal_2017, Herman_2022}. In this work, we explicitly assume that the spinning dust contribution to the polarized emission is zero.

\section{Material Properties of Interstellar Grains}
\label{sec:materials}

\subsection{Astrodust}
\label{subsec:astrodust}

If a single material is representative of the bulk of the interstellar dust mass, then it must contain most of the elements observed to be depleted from the gas phase, notably C, O, Mg, Si, and Fe \citep[e.g.,][]{Jenkins_2009}. We posit that interstellar grains consist of different materials on very small ($\lesssim 50$\,\AA) scales, but are mixed and (roughly) compositionally uniform on $\gtrsim 500$\,\AA\ scales. The composite grain consists then of a large collection of nano-scale domains, including voids, such that any two grains of sufficient size would have approximately equal amounts of each material and thus comparable optical properties. We term this composite material ``astrodust.'' We recently presented a detailed accounting of the putative composition of astrodust \citep{Draine+Hensley_2021a}, which we summarize briefly below.

As amorphous silicates are well-attested in the diffuse ISM \citep[e.g.,][]{vanBreemen+etal_2011}, we assume some of the grain material to be in the form of silicates with nominal composition Mg$_{1.3}$(Fe,Ni)$_{0.3}$SiO$_{3.6}$. Fe is observed to be highly depleted from the gas phase, but our adopted silicate composition accounts for only $\simeq 25$\% of the solid phase Fe. The rest is assumed to be in the form of Fe oxides, carbides, and sulfur compounds. We do not consider the possibility of ferromagnetic Fe nanoparticles in this work. The PAH component employed in this work accounts for less than half of the solid phase carbon, and so the rest is assumed to be in astrodust. We posit that 3\,ppm of C is in CaCO$_3$ to account for the solid phase Ca abundance and that the remainder is in the form of hydrocarbons. The remaining solid phase abundances of Si and Al are accounted for with SiO$_2$ and Al$_2$O$_3$, respectively. Finally, we assume a grain porosity $\mathcal{P} = 0.2$. As demonstrated by \citet{Draine+Hensley_2021a}, this model accounts for the solid phase abundances of all of the principal dust-forming elements with the exception of O, which remains a longstanding issue \cite[see discussion in][]{Jenkins_2009}. With this assumed composition and relative solid phase abundances following \citet{Hensley+Draine_2021}, the astrodust mass density is $\rho_{\rm Ad} = 3.42\times\left(1-\mathcal{P}\right) = 2.74$\,g\,cm$^{-3}$.

We have recently demonstrated that an astrodust-based model employing oblate grains with $\mathcal{P} \lesssim 0.75$ can reproduce the observed ratio between polarized FIR emission and polarized optical extinction \citep{Draine+Hensley_2021c}. For $\mathcal{P} \simeq 0.2$, the axial ratio is constrained to be $\gtrsim 1.4$. Here we assume all astrodust grains are oblate with an axial ratio of 1.4:1. In reality, we expect there to exist a distribution of grain shapes, likely dependent on grain size. \citet{Draine+Hensley_2021b} demonstrated that, for $\lambda \gg a$, polarization cross sections of grains averaged over a continuous distribution of ellipsoids (CDE) are well-approximated by 2:1 oblate spheroids for the CDEs proposed by \citet{Ossenkopf+Henning+Mathis_1992} and \citet{Zubko+etal_1996}. Thus the assumption of a single grain shape is a reasonable approximation for our present purposes.

Deriving the astrodust size distribution $dn_{\rm Ad}/da$ is a principal goal of this work. The only physical constraint we impose is to truncate the size distribution below a grain radius of 4.5\,\AA, where silicate grains are estimated to be photolytically unstable in the interstellar radiation field \citep{Guhathakurta+Draine_1989}. Given a size distribution, the total grain volume per H of composition $i$ is

\begin{equation} \label{eq:vol}
    V_i = \int da \frac{4}{3}\pi a^3 \left(\frac{1}{n_{\rm H}} \frac{dn_i}{da}\right)
    ~~~.
\end{equation}
The model is calibrated such that the solid phase abundances in Table~2 of \citet{Hensley+Draine_2021} are accounted for, with the exceptions of C and O, when $V_{\rm Ad} = 5.58\times10^{-27}$\,cm$^3$\,H$^{-1}$ \citep[see][Table~2, for $\mathcal{P} = 0.2$]{Draine+Hensley_2021a}.

We employ the dielectric function presented in \citet{Draine+Hensley_2021a} to calculate cross sections for all astrodust grains. Nanometer-sized grains are of comparable size to domains in our idealized astrodust grains and should more appropriately be modeled with dielectric functions specific to each material. We defer these refinements to future work, particularly as the optical properties of these very small grains are not well constrained by present data.

When computing spinning dust emission from astrodust nanoparticles, we assume the intrinsic electric dipole moment $\mu$ of a grain scales with the number of atoms $N_{\rm at}$ as \citep{Draine+Lazarian_1998a}

\begin{equation} \label{eq:beta_spdust}
    \mu \approx \beta_{\rm ed} N_{\rm at}^{1/2}
    ~~~.
\end{equation}
We assume that these nano-grains have a characteristic composition of Mg$_{1.3}$Fe$_{0.3}$SiO$_{3.6}$ and so have a number of atoms

\begin{equation}
    N_{\rm at}^{\rm Ad} = 1 + \left\lfloor 397\left(1-\mathcal{P}\right)\left(\frac{a}{10\,\text{\AA}}\right)^3\right\rfloor
    ~~~,
\end{equation}
where $\left\lfloor\,\right\rfloor$ is the floor function. We adopt $\beta_{\rm ed} = 0.3$\,D for astrodust, consistent with proposed spinning nanosilicate models of the AME \citep{Hoang+Vinh+QuynhLan_2016, Hensley+Draine_2017b} and within the range derived from quantum mechanical calculations of dipole moments of silicate nanoclusters \citep{MaciaEscatllar_2020}. In practice, our best fit model has only a small population of astrodust nanoparticles and so is insensitive to these details.

\subsection{PAHs}
The PAH model adopted in this work follows \citet{Draine_2021}. We briefly summarize the relevant aspects here.

The mass density of PAHs is taken to be $\rho_{\rm PAH} = 2.0$\,g\,cm$^{-3}$. This differs from some earlier works \citep[e.g.,][]{Draine+Lee_1984,Draine+Li_2007} that used the bulk graphite density of 2.2\,g\,cm$^{-3}$.

The adopted cross sections for PAHs are based on those presented in \citet{Draine+Li_2007}. As in that study, we combine these cross sections with those of graphite to account for continuum opacity between the prominent MIR features. Following \citet{Draine_2021}, we employ the cross sections for ``turbostratic'' graphite, a material consisting of randomly oriented graphitic microcrystals, derived by \citet{Draine_2016}. We assume the PAH cross sections depend only upon the grain effective radius and grain charge. For the latter, we consider a binary charge parameter, i.e., a PAH is either neutral or ionized.

Let $C^{\rm neu}\left(a,\lambda\right)$ and $C^{\rm ion}\left(a,\lambda\right)$ denote the ``pure PAH'' cross sections for neutral and ionized PAHs, respectively, and $C^{\rm gra}\left(a,\lambda\right)$ denote the cross section for graphite. Then the total cross section in our model is

\begin{equation} \label{eq:pah_graphite}
    C^{{\rm PAH}, Z}\left(a,\lambda\right) = \left[1-\xi\left(a\right)\right]C^Z\left(a,\lambda\right) + \xi\left(a\right) C^{\rm gra}\left(a,\lambda\right)
    ~~~,
\end{equation}
where $Z$ denotes either ionized (``ion'') or neutral (``neu'') PAHs and $\xi\left(a\right)$ is given by

\begin{align}
    \xi\left(a\right) &= 0.01, \quad a \leq 50\,\text{\AA} \\
    &= 0.01 + 0.99\left[1 - \left(\frac{50\,\text{\AA}}{a}\right)^3\right], \quad a > 50\,\text{\AA}
    ~~~.
\end{align}

In this work we adopt the functional form of the PAH size distribution employed in \citet{Draine_2021}, viz.,

\begin{equation} \label{eq:dnda_pah}
    \frac{1}{n_{\rm H}} \frac{dn_{\rm PAH}}{da} = \sum_{j=1}^2 \frac{B_j}{a} {\rm exp}\left\{-\frac{\left[\ln \left(a/a_{0,j}\right)\right]^2}{2\sigma^2}\right\}
    ~~~,
\end{equation}
for $a > a_{\rm min, PAH}$ and zero otherwise. As smaller carbonaceous grains are photolytically unstable in the interstellar radiation field, we adopt $a_{\rm min, PAH} = 4.0$\,\AA\ \citep{Guhathakurta+Draine_1989, Draine_2021}. We adopt values of $a_{0,1} = 4.0$\,\AA, $a_{0,2} = 30\,$\AA, and $\sigma = 0.40$. The $B_j$ factors are given by

\begin{align}
    B_j &= \frac{3e^{-4.5\sigma^2}}{\left(2\pi\right)^{3/2}\rho_{\rm PAH} a_{0,j}^3\sigma} \times \nonumber \\
    &\frac{b_{{\rm C},j} m_{\rm C}}{1 + {\rm erf}\left[3\sigma/\sqrt{2} + \ln\left(a_{0,j}/a_{\rm min,PAH}\right)/\left(\sigma\sqrt{2}\right)\right]}
    ~~~,
\end{align}
where $b_{{\rm C},j}$ is the C abundance per H in the $j$th lognormal component, $m_{\rm C}$ is the mass of a carbon atom, and erf is the error function. We leave $B_1$ and $B_2$ as parameters to be fit.

To partition the PAH size distribution into the neutral and ionized PAH components, we use the analytic function for the PAH ionization fraction $f_{\rm ion}\left(a\right)$ corresponding to the ``standard'' ionization model in \citet{Draine_2021}:

\begin{equation} \label{eq:fion}
    f_{\rm ion}\left(a\right) = 1 - \frac{1}{1 + a/10\,\text{\AA}}
    ~~~.
\end{equation}

The above model is then employed to translate our adopted cross sections for neutral PAHs, ionized PAHs, and graphite to wavelength-dependent extinction and emission per H atom from PAHs. We assume that the PAH component of our model does not contribute to polarized extinction or emission.

While the representative sample of molecules considered by \citet{Draine+Lazarian_1998b} has $0.14 < \beta_{\rm ed}/{\rm D} < 1.52$, the dipole moment distribution of interstellar grains is largely unconstrained. Given the large observational and theoretical uncertainties, we elect not to fit a parametric model for the electric dipole moment distribution in our global optimization. Rather, we assume that 35\% of PAHs have $\beta_{\rm ed} = 0.04$\,D, 4\% have $\beta_{\rm ed} = 0.3$\,D, and the rest have no dipole moment. We use these post hoc values to demonstrate compatibility of our best fit PAH size distribution with the observed level and frequency spectrum of AME. This is discussed in further in Section~\ref{subsec:irem}.

We compute the number of atoms in a PAH as

\begin{equation}
    N_{\rm at}^{\rm PAH} = N_{\rm C}^{\rm PAH} + N_{\rm H}^{\rm PAH}
    ~~~,
\end{equation}
where the number of carbon atoms is

\begin{equation}
    N_{\rm C}^{\rm PAH} = 1 + \left\lfloor 417\left(\frac{a}{10\,\text{\AA}}\right)^3 \right\rfloor
\end{equation}
and the number of hydrogen atoms is \citep[][Equation~8]{Draine+Li_2001}

\begin{align}
    N_{\rm H}^{\rm PAH} = &\begin{cases}
\left\lfloor\frac{1}{2}N_{\rm C}^{\rm PAH} + \frac{1}{2}\right\rfloor
& N_{\rm C}^{\rm PAH}\ < 25 \\
\left\lfloor\frac{5}{2}\sqrt{N_{\rm C}^{\rm PAH}} + \frac{1}{2}\right\rfloor
& 25\, \leq N_{\rm C}^{\rm PAH}\ < 100 \\
\left\lfloor\frac{1}{4}N_{\rm C}^{\rm PAH} + \frac{1}{2}\right\rfloor
& N_{\rm C}^{\rm PAH}\ \geq 100
~~~.
\end{cases}
\end{align}

\section{Data Model} \label{sec:data_model}

\subsection{Observational Constraints}
\label{subsec:observations}
The objective of this work is to construct a physical model of interstellar dust that reproduces the observed extinction, emission, and polarization from dust in the diffuse ISM while accounting for the solid phase abundances of elements observed to be depleted from the gas phase. Thus, we require the wavelength-dependent extinction, polarized extinction, emission, and polarized emission per H atom that typifies observations of the diffuse ISM. We rely on our recent determination of these quantities \citep{Hensley+Draine_2021}, which we briefly summarize below.

The adopted extinction curve $\tau_\lambda/N_{\rm H}$ is based on a synthesis of the reddening laws of \citet{Cardelli+Clayton+Mathis_1989} (FUV), \citet{Gordon+etal_2009} (FUV), \citet{Schlafly+etal_2016} (optical and NIR), \citet{Fitzpatrick+etal_2019} (UV and optical), and \citet{Hensley+Draine_2020} (MIR) in the indicated ranges, and adopting $A_H/A_{K_s} = 1.55$ \citep{Indebetouw+etal_2005} to convert to total extinction. The extinction curve is then normalized to $N_{\rm H}/E(B-V) = 8.8\times10^{21}$\,cm$^{-2}$\,mag$^{-1}$ \citep{Lenz+Hensley+Dore_2017}. 

From 0.13--1.38\,$\mu$m, the adopted polarized extinction law $p_\lambda/N_{\rm H}$ follows a Serkowski law \citep{Serkowski_1973}

\begin{equation}
    p_\lambda = p_{\rm max} {\rm exp}\left[-K\ln^2\left(\lambda_{\rm max}/\lambda\right)\right]
    ~~~,
\end{equation}
with parameters $K = 0.87$ and $\lambda_{\rm max} = 0.55$\,$\mu$m \citep{Whittet_2003} and continues to longer wavelengths as a power law with $p_\lambda \propto \lambda^{-1.6}$ for $1.38 < \lambda/\mu{\rm m} < 4$ \citep{Martin+etal_1992}. The full curve is normalized to $p_V/E(B-V) = 0.13$\,mag$^{-1}$ \citep{Panopoulou+etal_2019,Planck_2018_XII}.

The adopted dust emission spectrum $\lambda I_\lambda/N_{\rm H}$ is based on the composite spectrum of \citet{Lai_2020} from 3--12\,$\mu$m and the spectrum of Dcld\,300.2--16.9 (B) from 12--38\,$\mu$m \citep{Ingalls+etal_2011}, with the synthesized spectrum normalized to the hydrogen column using the mean dust emission per $N_{\rm H}$ in the DIRBE bands \citep{Dwek+etal_1997}. From FIR to microwave wavelengths, we adopt the dust SEDs per $N_{\rm H}$ derived by \citet{Planck_Int_XVII} and \citet{Planck_Int_XXII}, with a 1.5\% correction to the 353\,GHz intensity following \citet{Planck_2018_XI}. The adopted dust emission spectrum in polarization $\left(\lambda P_\lambda/N_{\rm H}\right)_{\rm max}$ is based on the dust $E$-mode polarization spectrum derived by \citet{Planck_2018_XI}, where the ``max'' subscript indicates the intrinsic maximum polarization corresponding to the magnetic field lying in the plane of the sky. This spectrum has $p_{353}/(p_V/\tau_V) = 4.31$ \citep{Planck_2018_XII} using the adopted extinction and polarized extinction laws, where $p_{353}$ is the dust polarization fraction at 353\,GHz and $\tau_V$ is the dust optical depth in the $V$ band.

Recently, \citet{Gordon_2021} made a new determination of the MIR extinction law of the diffuse ISM using Spitzer observations of moderately reddened sources. Their curve is significantly steeper from $\sim3$--6\,$\mu$m than the extinction curve of \citet{Hensley+Draine_2020} based on the sightline toward Cyg\,OB2-12 that we adopt here. The steepness of the interstellar extinction law in this wavelength range is particularly uncertain: some studies have also found evidence for a steeper extinction law \citep[e.g.,][]{Chen+etal_2018,Wang+Chen_2019}, while others are consistent with our fiducial extinction law \citep[e.g.,][]{Indebetouw+etal_2005,Shao+etal_2018}. More work is required to clarify this discrepancy. 

Reproducing the greater MIR extinction adopted here is more challenging with respect to interstellar abundance constraints than reproducing the steeper curve. Indeed, some authors have modified previous dust models with the addition of large graphitic grains to account for relatively strong MIR extinction \citep{Wang+Li+Jiang_2015}. The adopted dielectric function of astrodust is (by design) sufficiently absorptive in the $5$--$8\,\mu$m range to be consistent with both the strength and shape of the MIR extinction curve of \citet{Hensley+Draine_2021} without the need to invoke very large grains. Should further observations determine that the MIR extinction curve is steeper from 3--6\,$\mu$m than we employ here, we anticipate that the astrodust dielectric function could be straightforwardly modified following the methods of \citet{Draine+Hensley_2021a} to accommodate this difference while still respecting interstellar abundance constraints as well as observations of dust extinction, emission, and polarization at other wavelengths.

\subsection{Fitting Framework}
\label{subsec:fitting}

The adopted observational constraints discussed in the previous section provide smooth curves describing:

\begin{itemize}
\item the total extinction per H atom $\tau_\lambda/N_{\rm H}$ from 0.1 to 30\,$\mu$m; 
\item the maximum polarized extinction per H atom $p_\lambda/N_{\rm H}$ from 0.13 to 4\,$\mu$m; 
\item the total emission per H atom $I_\lambda/N_{\rm H}$ from 3 to 10$^4$\,$\mu$m; 
\item and the maximum polarized emission per H atom $P_\lambda/N_{\rm H}$ from 850 to 6800\,$\mu$m. 
\end{itemize}
We discretize these into bins of 100 points per decade in $\lambda$. This yields 248 points in extinction, 153 in polarized extinction, 353 in emission, and 91 in polarized emission.

As discussed in Section~\ref{sec:modeling}, the dust model parameters to be constrained are the astrodust size distribution $dn_{\rm Ad}/da$, the astrodust alignment function $f_{\rm align}^{\rm Ad}$, and the PAH size distribution parameters $B_1$ and $B_2$. While some previous studies have discretized the size distributions and alignment functions into a large number of size bins each fit as a free parameter subject to smoothness constraints \citep{Kim+Martin_1995, Draine+AllafAkbari_2006, Draine+Fraisse_2009}, we opt instead to construct simple parametric representations and thereby fit a much smaller set of parameters.

The astrodust size distribution for 4.5\,\AA\ $< a < 5\,\mu$m is modeled as

\begin{align} \label{eq:dnda}
    \frac{1}{n_{\rm H}} &\frac{dn_{\rm Ad}}{da}\left(a\right) =  \frac{B_{\rm Ad}}{a} {\rm exp} \left\{ -\frac{ \left[\ln\left(a/a_{0,{\rm Ad}}\right)\right]^2}{2\sigma^2_{\rm Ad}} \right\} + \nonumber \\
    &\frac{A_0}{a}~{\rm exp}\left\{\sum_{i=1}^5 A_i \left[\ln\left(\frac{a}{\text{\AA}}\right)\right]^i\right\}
\end{align}
and set to zero for sizes outside this range. Here $B_{\rm Ad}$, $a_{0,{\rm Ad}}$, $\sigma_{\rm Ad}$, and $A_i$ for $i \in [0,5]$ are scalar fitting parameters. Thus, we fit the astrodust size distribution with a total of nine free parameters. We do not explore size distributions having $\sigma_{\rm Ad} < 0.25$ or $\sigma_{\rm Ad} > 0.8$, but otherwise do not impose any restrictions on parameter values. 

The functional form of the size distribution is selected for its relative simplicity and smoothness while still being able to generate size distributions with significant structure like those found in previous studies \citep[e.g.,][]{Kim+Martin_1995, Draine+Fraisse_2009}. The log-normal component is included as a parallel to the log-normal components used in the PAH size distribution. As detailed in Section~\ref{subsec:ext}, we find the astrodust grains accounted for in the log-normal component to be important for the UV extinction.

The astrodust alignment function is modeled as

\begin{equation} \label{eq:falign}
    f_{\rm align}^{\rm Ad}\left(a\right) = \frac{f_{\rm max}}{1 + \left(a_{\rm align}/a\right)^{\alpha_{\rm align}}}
    ~~~,
\end{equation}
where $a_{\rm align}$, $\alpha_{\rm align} > 0$, and $f_{\rm max} \leq 1$ are scalar parameters to be fit. This simple functional form---transitioning from no alignment at small grain sizes, to $f_{\rm align} = f_{\rm max}/2$ for $a = a_{\rm align}$, to an alignment fraction $f_{\rm max}$ at large grain sizes---is compatible with alignment functions that have been inferred from observations \citep[e.g.,][]{Kim+Martin_1995, Draine+Fraisse_2009, Guillet+etal_2018} as well as theoretical studies of grain alignment via radiative torques \citep[e.g.,][]{Hoang+Lazarian_2016b}.

We quantify the distance of the resulting extinction, polarized extinction, emission, and polarized emission from the adopted observational constraints by computing the sum of squares of the normalized residuals $\Delta$, i.e.,

\begin{equation} \label{eq:Delta}
    \Delta = \sum \left(\frac{{\rm Data}-{\rm Model}}{\rm Data}\right)^2
    ~~~,
\end{equation}
where ``Data'' refers to our adopted observational constraints in extinction, polarized extinction, emission, and polarized emission. As the wavelength dependence of dust extinction, emission, and polarization is relatively well-constrained over the specified wavelength ranges and we wish for the model to reproduce all observed features therein, Equation~\eqref{eq:Delta} applies a uniform weighting in $\log\left(\lambda\right)$. Finally, we use a Nelder-Mead minimization routine \citep{Nelder+Mead_1965, SciPy} to find the parameter values yielding the size distribution and alignment function that minimize $\Delta$.

The principal goal of this work is construction of a grain model based on physically realistic material properties that can reproduce the fundamental observables of dust in the diffuse ISM. In Section~\ref{subsubsec:stats} we present a roadmap for developing the present analysis into a full statistical inference of dust properties.

\section{Results}
\label{sec:results}

In this section, we present our new model of interstellar dust based on astrodust and PAHs. We provide an overview of the size distributions, alignment functions, and overall fit to the observational data in Section~\ref{subsec:results_overview} before describing in greater detail the ability of the model to reproduce the observed extinction (Section~\ref{subsec:ext}), polarized extinction (Section~\ref{subsec:extpol}), emission (Section~\ref{subsec:irem}), polarized emission (Section~\ref{subsec:ipol}), and scattering (Section~\ref{subsec:scatt}).

\subsection{Overview}
\label{subsec:results_overview}

\begin{deluxetable}{ccc}
  \tablewidth{0pc}
      \tablecaption{Best Fit Parameter Values\label{table:parameters}}
    \tablehead{\colhead{Parameter} & \colhead{Value} & \colhead{Equation}}
    \startdata
    $B_1$ & $7.52\times10^{-7}$\,H$^{-1}$ & \ref{eq:dnda_pah} \\
    $B_2$ & $8.09\times10^{-10}$\,H$^{-1}$ & \ref{eq:dnda_pah} \\
    $B_{\rm Ad}$ & $3.31\times10^{-10}$\,H$^{-1}$ & \ref{eq:dnda} \\
    $a_{0,{\rm Ad}}$ & 63.8\,\AA & \ref{eq:dnda} \\
    $\sigma_{\rm Ad}$ & 0.353 & \ref{eq:dnda} \\
    $A_0$ & $2.97\times10^{-5}$\,H$^{-1}$ & \ref{eq:dnda} \\
    $A_1$ & -3.40 & \ref{eq:dnda} \\
    $A_2$ & -0.807 & \ref{eq:dnda} \\
    $A_3$ & 0.157 & \ref{eq:dnda} \\
    $A_4$ & $7.96\times10^{-3}$ & \ref{eq:dnda} \\
    $A_5$ & $-1.68\times10^{-3}$ & \ref{eq:dnda} \\
    $a_{\rm align}$ & 0.0749\,$\mu$m & \ref{eq:falign} \\
    $\alpha_{\rm align}$ & 1.80 & \ref{eq:falign} \\
    $f_{\rm max}$ & 1.00 & \ref{eq:falign} \\
    \enddata
\end{deluxetable}

\begin{deluxetable*}{cc}
  \tablewidth{0pc}
      \tablecaption{Derived Model Quantities\label{table:quantities}}
    \tablehead{\colhead{Quantity} & \colhead{Value}}
    \startdata
    $V_{\rm Ad}$ & $3.92\times10^{-27}$\,cm$^{3}$\,H$^{-1}$ \\
    $V_{\rm PAH}$ & $5.51\times10^{-28}$\,cm$^{3}$\,H$^{-1}$ \\
    $M_{\rm Ad}/M_{\rm H}$ & 0.0064 \\
    $M_{\rm PAH}/M_{\rm H}$ & 0.0007 \\
    $M_{\rm d}/M_{\rm H} \equiv \left(M_{\rm Ad} + M_{\rm PAH}\right)/M_{\rm H}$ & 0.0071 \\
    $M_{\rm PAH}/M_{\rm d}$ & 0.093 \\
    $q_{\rm PAH}$ & 5.91\% \\
    $\langle f_{\rm align}^{\rm Ad}\rangle$ & 0.70 \\
    $\Sigma_{\rm Ad}$ & $3.00\times10^{-21}$\,cm$^2$\,H$^{-1}$ \\
    $\Sigma_{\rm PAH}$ & $1.74\times10^{-20}$\,cm$^2$\,H$^{-1}$ \\
    $A_{0.55\,\mu{\rm m}}/N_{\rm H}$ & $3.24\times10^{-22}$\,mag\,cm$^2$ \\
    $N_{\rm H}/\left(A_{0.44\,\mu{\rm m}}-A_{0.55\,\mu{\rm m}}\right)$ & $9.47\times10^{21}$\,cm$^{-2}$\,mag$^{-1}$\\
    $R\left(55\right) \equiv A_{0.55\,\mu{\rm m}} /\left(A_{0.44\,\mu{\rm m}}-A_{0.55\,\mu{\rm m}}\right)$ & 3.07 \\
    $\left[p_{0.55\,\mu{\rm m}}/\left(A_{0.44\,\mu{\rm m}}-A_{0.55\,\mu{\rm m}}\right)\right]^{\rm max}$ & 0.143\,mag$^{-1}$ \\
    $\left(P_{353}/I_{353}\right)^{\rm max}$ & 19.2\% \\
    $R_{P/p} \equiv P_{353}/p_{0.55\,\mu{\rm m}}$ & 4.54\,MJy\,sr$^{-1}$ \\
    $R_{S/V} \equiv \left(\tau_{0.55\,\mu{\rm m}}/I_{353}\right) R_{P/p}$ & 3.78 \\
    $\mathcal{R}/A_{0.55\,\mu{\rm m}} = \int_{50\,\mu{\rm m}}^{1\,{\rm mm}} I_\lambda d\lambda / A_{0.55\,\mu{\rm m}}$ & $8.30\times10^{-7}$\,W\,m$^{-2}$\,sr$^{-1}$\,mag$^{-1}$ \\
    \enddata
\end{deluxetable*}

\begin{figure}
    \centering
        \includegraphics[width=\columnwidth]{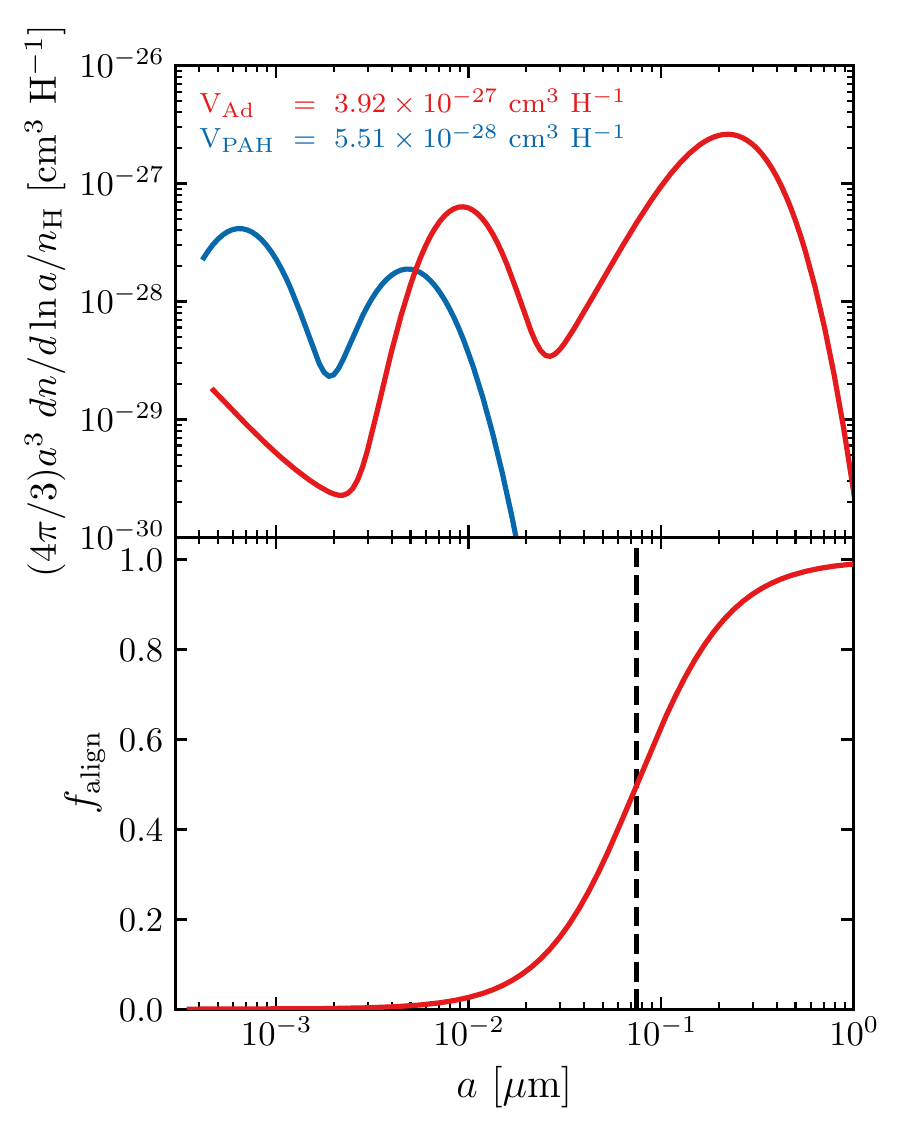}
    \caption{The size distributions of the astrodust (red) and PAH (blue) components of the model are shown in the top panel. The best fit astrodust alignment function is presented in the bottom panel.} \label{fig:model_dnda} 
\end{figure}

\begin{figure*}
    \centering
        \includegraphics[width=\textwidth]{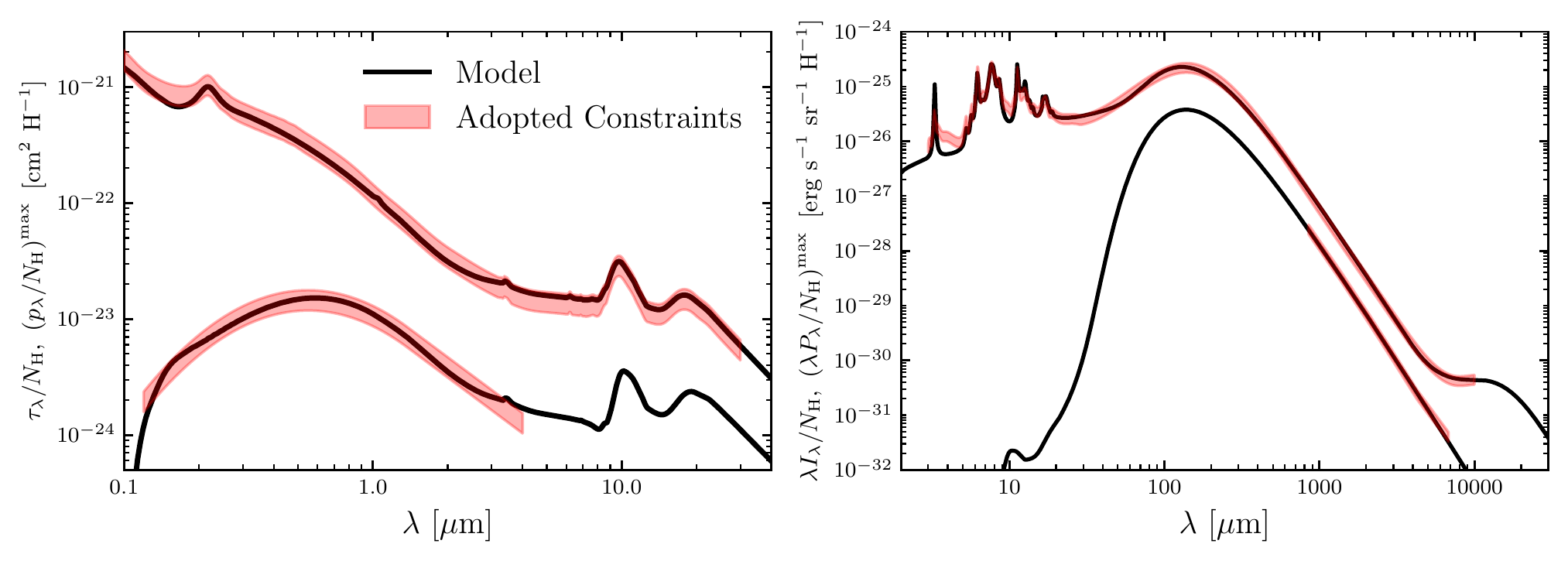}
    \caption{The left panel illustrates the total and polarized extinction arising from the model with size distributions and alignment function shown in Figure~\ref{fig:model_dnda}. These are compared to the adopted constraints on the total and polarized extinction (red shaded regions, corresponding to the adopted constraint $\pm20$\%). The right panel illustrates the total and polarized emission from this model. As with extinction, the adopted observational constraints are plotted as a red shaded region representing deviations of $\pm20$\%. The model fits the data to within 20\% at most wavelengths.} \label{fig:model_summary}
\end{figure*}

\begin{figure*}
    \centering
         \includegraphics[width=\textwidth]{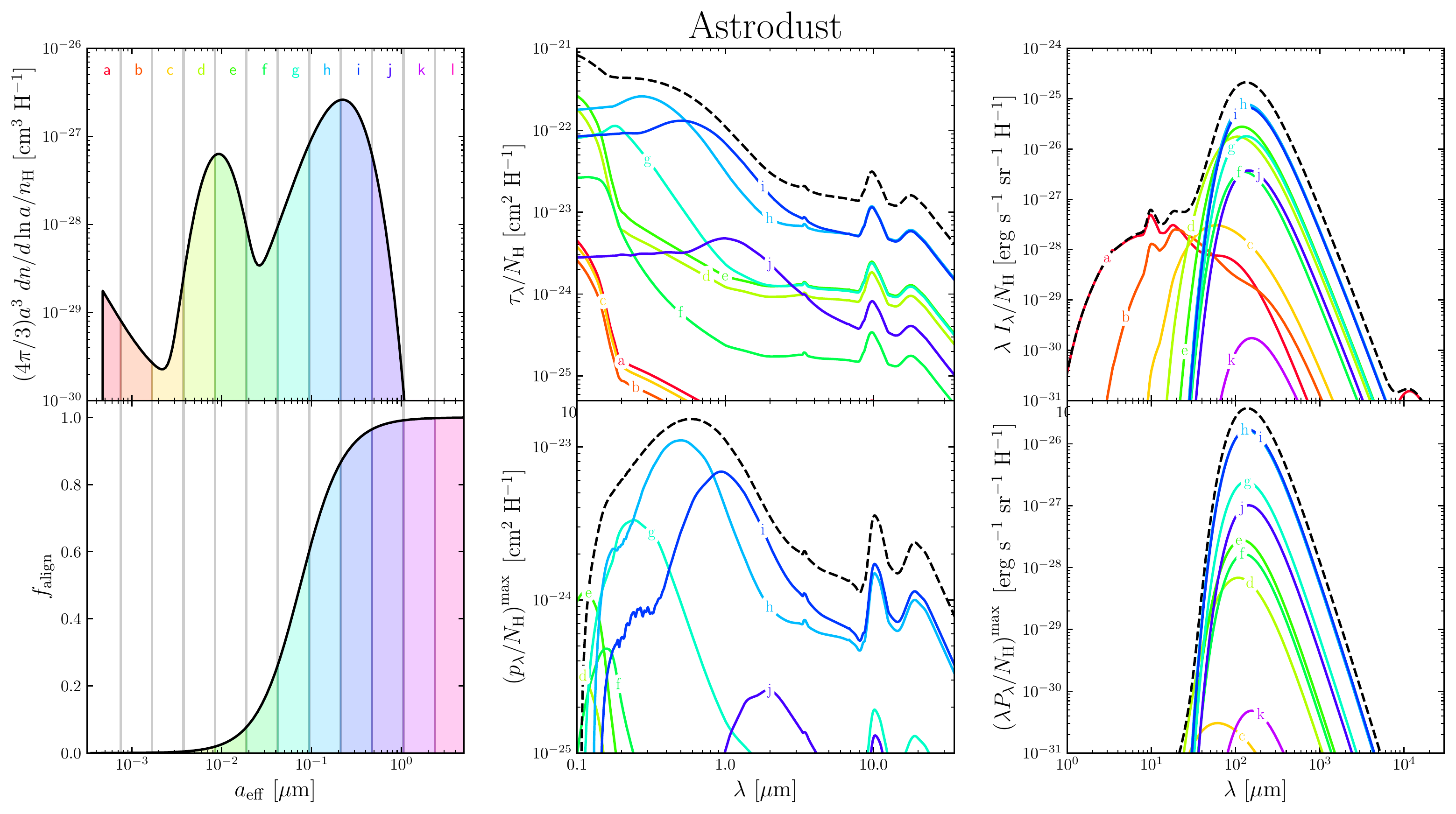}
        \includegraphics[width=\textwidth]{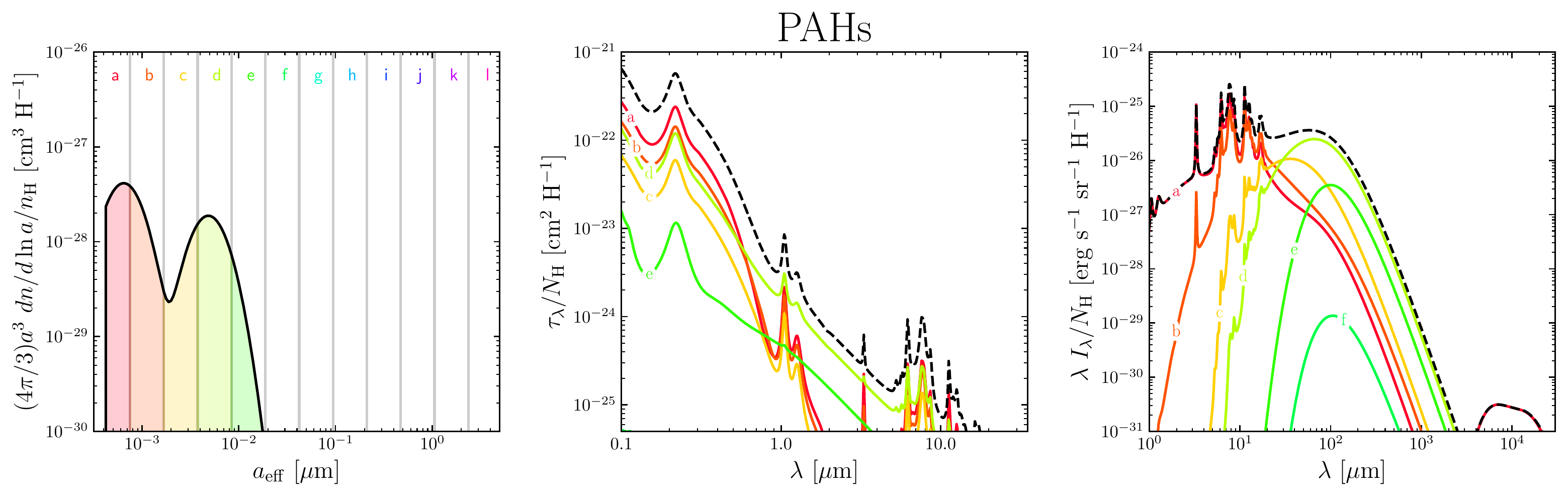}
    \caption{We divide the size distributions of astrodust grains (top left) and PAHs (bottom left) into discrete bins, each assigned a different color (note that the size binning and color scheme are the same for both compositions). We then plot the extinction (middle) and emission (right) arising from all grains within each size bin. For astrodust grains only, the alignment function (left), polarized extinction (middle), and polarized emission (right) are presented in the second row following the same color scheme---in our model, PAHs are assumed to produce no polarization. In the middle and right panels of each row, we plot the sum over all size bins with a black dashed line.} \label{fig:size_analysis} 
\end{figure*}

The best fit astrodust and PAH size distributions are presented in the top panel of Figure~\ref{fig:model_dnda}, while the bottom panel of Figure~\ref{fig:model_dnda} presents the derived astrodust alignment function. The best fit parameters of each are given in Table~\ref{table:parameters}.

The total astrodust volume in the model is $V_{\rm Ad} = 3.92\times10^{-27}$\,cm$^3$\,H$^{-1}$, 70\% of the estimated $5.58\times10^{-27}$\,cm$^3$\,H$^{-1}$ based on interstellar abundances (see Section~\ref{subsec:astrodust}). The observed solid phase abundances are thus able to comfortably accommodate the fit size distribution. With these astrodust and PAH size distributions, the total dust mass fraction in PAHs containing fewer than 1000 C atoms $q_{\rm PAH} = 5.9$\%, somewhat higher than the 4.6\% of the \citet{Draine+Li_2007} model or the 3.8\% of \citet{Draine_2021}. The fraction of the total dust mass in all PAHs is 9.3\%. The $B_1$ and $B_2$ parameters governing the PAH mass in the first and second log-normal components are factors of 1.23 and 2.60 larger, respectively, than those employed by \citet{Draine_2021}. The total dust to gas mass ratio $M_{\rm d}/M_{\rm H}$ of the model is 0.0071. While astrodust dominates the dust mass, PAHs in the model have a effective surface area\footnote{Equation~\eqref{eq:Sigma} is inexact for spheroids but underestimates the surface area of a 1.4:1 oblate spheroid by only 2\%.} per H atom $\Sigma_{\rm PAH}$ nearly six times larger than $\Sigma_{\rm Ad}$, where

\begin{equation} \label{eq:Sigma}
\Sigma_i \equiv \int da\, 4\pi a^2 \left(\frac{1}{n_{\rm H}}\frac{dn_i}{da}\right)
~~~.
\end{equation}
These and other derived properties of the model are summarized in Table~\ref{table:quantities}.

The astrodust size distribution shares features with previous size distributions based on (astro)silicates, such as a peak near 0.2\,$\mu$m, a rapid decline to larger grain sizes, and numerous grains at smaller grain sizes \citep[e.g.,][]{Weingartner+Draine_2001, Draine+Fraisse_2009}. The bimodal nature of the astrodust size distribution at $a > 30$\,\AA\ is also seen in some earlier studies that derived size distribution constraints based on polarized extinction \citep[e.g.,][]{Kim+Martin_1995, Draine+Fraisse_2009}. \citet{Kim+Martin_1995} in particular note that a deficit of grains between the two maxima may be an artifact of the adopted optical constants that leads to features in the polarized extinction curve for grains in this size range. Since the astrodust dielectric function is largely based on that of astrosilicate at optical wavelengths, it is perhaps unsurprising that a similar conclusion is reached here. We explore this further in our discussion of the polarized extinction in Section~\ref{subsec:extpol}. However, multi-modal size distributions have been derived from models of grain injection, destruction, and growth processes and so need not be considered unphysical \citep[e.g.,][]{Li2021,Hirashita2022}.

The astrodust size distribution in Figure~\ref{fig:model_dnda} has a small population of nanoparticles with $a < 20$\,\AA\ that accounts for 0.3\% of the total astrodust mass. The size distribution in this regime is poorly constrained by current data, but as we discuss in the following sections these grains do slightly improve the global fit through modest contributions to the MIR and microwave emission.

The best fit astrodust alignment function suggests grains attain a fractional alignment of 1/2 at a grain size of 0.075\,$\mu$m and nearly perfect alignment at 1\,$\mu$m. These numerical values are broadly consistent with detailed models of grain alignment for grains with large enough magnetic susceptibility to undergo paramagnetic relaxation on timescales short compared to disalignment via gas collisions \citep{Hoang+Lazarian_2016b}. In total, 70\% of the astrodust grain mass is aligned, where

\begin{equation}
    \langle f_{\rm align}^{\rm Ad}\rangle \equiv \frac{1}{V_{\rm Ad}} \int da \frac{4}{3} \pi a^3 f_{\rm align}\left(a\right) \frac{1}{n_{\rm H}} \frac{dn_{\rm Ad}}{da}
    ~~~.
\end{equation}
This corresponds to 63\% of the total dust mass.

The resulting extinction, emission, and polarization are summarized in Figure~\ref{fig:model_summary}, where they are compared to the observational constraints outlined in Section~\ref{subsec:observations}. Overall, the model provides an excellent fit to the observational constraints, deviating by less than 20\% at most wavelengths and often considerably less. In the following sections, we discuss fits to the extinction, polarized extinction, emission, and polarized emission with greater scrutiny, highlighting in particular places where the modeling could be further improved.

To clarify which grain sizes and compositions are responsible for various features in the model extinction, emission, and polarization, Figure~\ref{fig:size_analysis} illustrates the contributions from discrete size distribution bins to the model curves in Figure~\ref{fig:model_summary}. We examine this figure in greater detail throughout the following sections.

\subsection{Extinction}
\label{subsec:ext}

\begin{figure}
    \centering
        \includegraphics[width=\columnwidth]{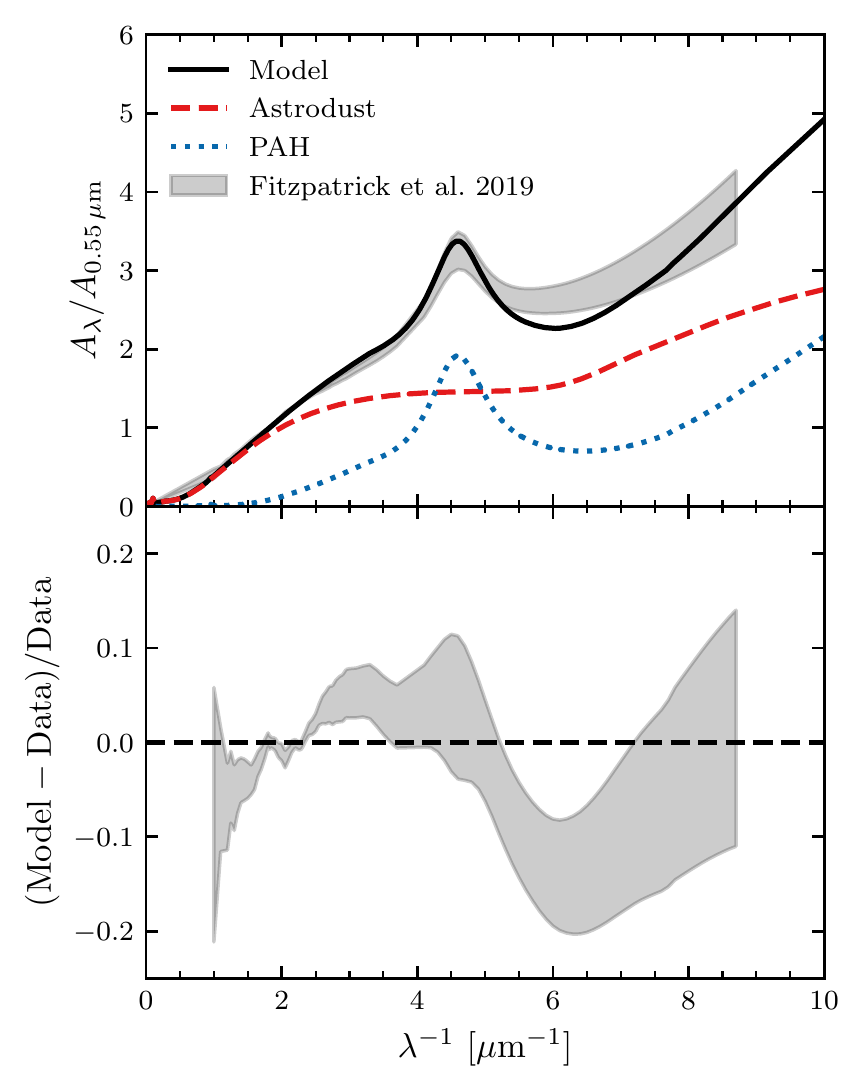}
    \caption{The top panel compares our model UV/optical extinction curve (black solid) to the determination of the mean Milky Way extinction curve by \citet{Fitzpatrick+etal_2019} (gray band representing the reported $1\sigma$ uncertainty). The total extinction is broken down into the contributions from astrodust (red dashed) and PAHs (blue dashed). The bottom panel quantifies the deviation of the model from the data. The agreement is within 20\% at all wavelengths, though there is a clear deficit of extinction from the model on the short wavelength side of the 2175\,\AA\ feature.} \label{fig:model_extopuv} 
\end{figure}

\begin{figure*}
    \centering
        \includegraphics[width=\textwidth]{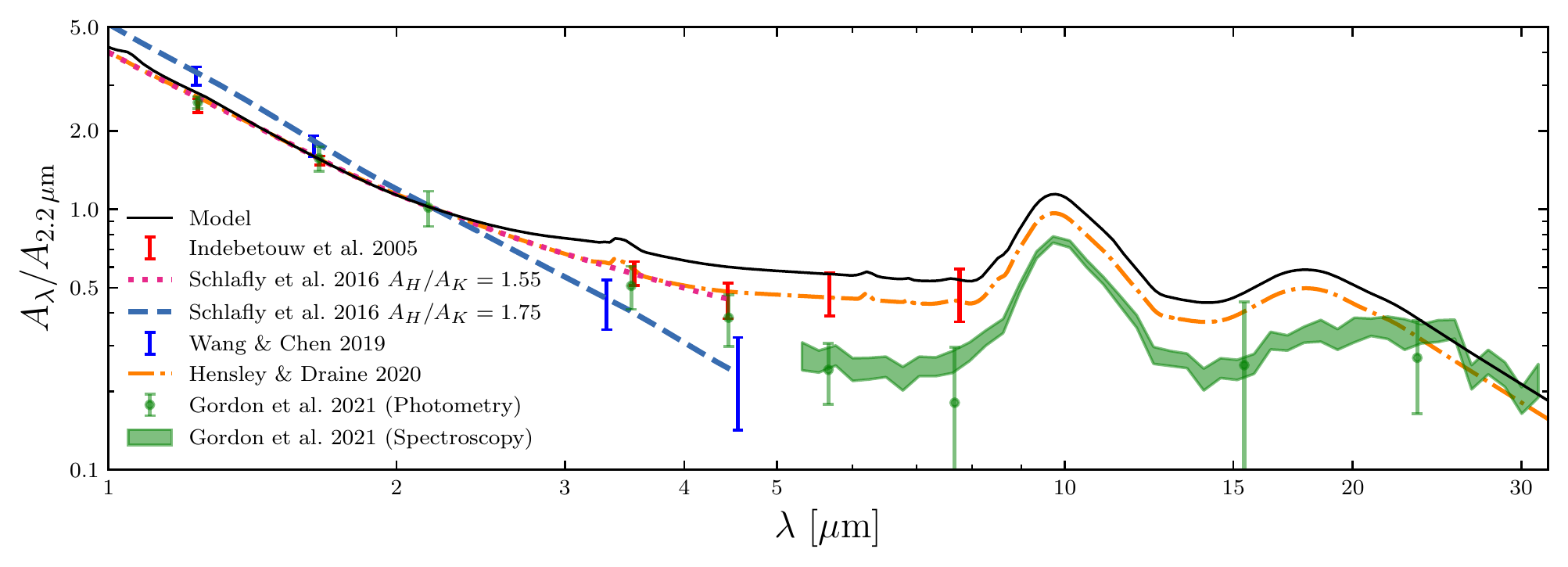}
    \caption{The total extinction from the model (black solid) is compared to various observational determinations representative of the diffuse ISM, normalized to unity at 2.2\,$\mu$m or K band \citep[cf.][Figures~3 and 4]{Hensley+Draine_2021}. These include both photometric \citep{Indebetouw+etal_2005, Schlafly+etal_2016, Wang+Chen_2019, Gordon_2021} and spectroscopic determinations \citep{Gordon_2021}, as well as a recent synthesis based on the NIR extinction curve of \citet{Schlafly+etal_2016} and spectroscopy toward Cyg~OB2-12 with ISO and Spitzer \citep{Hensley+Draine_2020}.} \label{fig:model_extir}
\end{figure*}

Recent studies of Galactic extinction employing spectrophotometry \citep[e.g.,][]{Fitzpatrick+etal_2019}, spectroscopy \citep[e.g.,][]{MaizApellaniz+etal_2014,Hensley+Draine_2020}, and photometry with modeling of bandpass effects \citep[e.g.,][]{Schlafly+etal_2016} permit model comparisons at monochromatic wavelengths. This avoids use of quantities like $E(B-V)$ and $R_V$ that involve the Johnson photometric bands.

\citet{Fitzpatrick+etal_2019} found that the mean Milky Way extinction curve has $R\left(55\right) \equiv A_{0.55\,\mu{\rm m}} /\left(A_{0.44\,\mu{\rm m}}-A_{0.55\,\mu{\rm m}}\right) = 3.02$, corresponding to $R_V = 3.10$. Our model curve has $R\left(55\right) = 3.07$, i.e., very close to the mean extinction law. The model has $A_{0.55\,\mu{\rm m}}/N_{\rm H} = 3.2\times10^{-22}\,$mag\,cm$^2$, in agreement with the adopted $A_V/N_{\rm H} = 3.5\times10^{-22}\,$mag\,cm$^2$ particularly in light of the $\sim10\%$ uncertainty in the latter \citep{Lenz+Hensley+Dore_2017, Nguyen+etal_2018}.

Figure~\ref{fig:model_extopuv} makes a detailed comparison between the mean Galactic extinction law derived by \citet{Fitzpatrick+etal_2019} and our model, both normalized to unity at 0.55\,$\mu$m. While agreement is very good from 1--3\,$\mu$m$^{-1}$, the UV extinction in the vicinity of the 2175\,\AA\ feature is lower in the model, with a maximum deficit of $14\%$ in the short wavelength wing of the feature. For $\lambda^{-1} > 6\,\mu$m$^{-1}$, grains in the log-normal component of the astrodust size distribution centered at 64\,\AA\ make an important contribution to the extinction (see Figure~\ref{fig:size_analysis}). As we discuss further in Section~\ref{subsec:extpol}, the deficit of grains of size $a \simeq 0.03\,\mu$m may be an artifact of the modeling of polarized extinction that limits our ability to obtain a better fit to the observed UV extinction.

The astrodust size distribution could be adjusted to improve agreement at these wavelengths, however such adjustments also affect the UV polarized extinction. As we discuss in Section~\ref{subsec:extpol}, modeling polarization at these wavelengths is particularly challenging given the sensitivity to grain shape---and thus our assumption of a single axial ratio versus a distribution of grain shapes---and limitations of the MPFA. We expect the simultaneous fit to both total and polarized extinction in the UV could be improved through a refined astrodust dielectric function, use of a distribution of grain shapes \citep[e.g., ``CDE2''][]{Ossenkopf+Henning+Mathis_1992, Draine+Hensley_2021b}, and full calculation of the orientation-dependent cross sections without invoking the MPFA. We defer these model refinements to future work.

Figure~\ref{fig:model_extir} presents the model IR extinction curve, normalized to unity at 2.2\,$\mu$m (roughly K band), as well as a number of observational determinations. Agreement between model and data is excellent from 1--2.2\,$\mu$m. At longer wavelengths, however, the model systematically overpredicts the extinction relative to K band. The significance of this overprediction is unclear given the level of observational uncertainty. The model is consistent with the extinction curve of \citet{Indebetouw+etal_2005} within the reported errors from 5--8\,$\mu$m but is $\sim30$\% higher from 3--5\,$\mu$m. In contrast, the model exceeds the spectroscopic extinction curve of \citet{Gordon_2021} by factors of 1.5--2 from 5.3--18\,$\mu$m.

The relatively flat MIR extinction law between 5--8\,$\mu$m derived by both \citet{Indebetouw+etal_2005} and \citet{Gordon_2021}, as well as other studies \citep[e.g.,][]{Lutz+etal_1996, Zasowski+etal_2009, Xue+etal_2016}, is in marked contrast with earlier determinations that featured a pronounced minimum at 8\,$\mu$m \citep[e.g.,][]{Rieke+Lebofsky_1985, Bertoldi+etal_1999}. Notably, the flat MIR extinction appears generic to the diffuse ISM and not just high $R_V$ lines of sight where previous models \citep[e.g.,][]{Weingartner+Draine_2001} have predicted significant flattening of the extinction law in the MIR. To account for this in the context of physical dust models, \citet{Wang+Li+Jiang_2015} invoked large graphitic grains to produce enough extinction without exceeding constraints on solid phase abundances. 

In contrast, our astrodust model demonstrates that the observed MIR extinction law can be accommodated by interstellar grains of typical sizes with a suitable dielectric function. We anticipate that MIR observations from the James Webb Space Telescope (JWST) will clarify the current observational discrepancies.

Our model makes predictions for extinction in PAH features, which has been observed on select lines of sight \citep{Schutte+etal_1998, Chiar+etal_2000, Chiar+etal_2013, Hensley+Draine_2020}. Visible in Figure~\ref{fig:model_extir} is a PAH feature at 1.05\,$\mu$m that, to our knowledge, has not been detected in either extinction or emission. A second, subdominant feature is also present at 1.26\,$\mu$m (see Figure~\ref{fig:size_analysis}). The presence of these features in the adopted PAH cross sections (dating to the work of \citet{Draine+Li_2007}) is based on laboratory measurements of weak electronic transitions in PAH cations \citep{Mattioda+etal_2005b}. With the best fit PAH size distribution, we find that the 1.05 and 1.26\,$\mu$m features have strengths $\Delta\tau/A_{0.55\,\mu{\rm m}} = 0.017$ and 0.004, respectively. Stringent constraints should be feasible with the NIRSpec instrument aboard JWST \citep{Jakobsen_2022}, and may already be possible with archival NIR data.
  
\subsection{Polarized Extinction}
\label{subsec:extpol}

\begin{figure}
    \centering
        \includegraphics[width=\columnwidth]{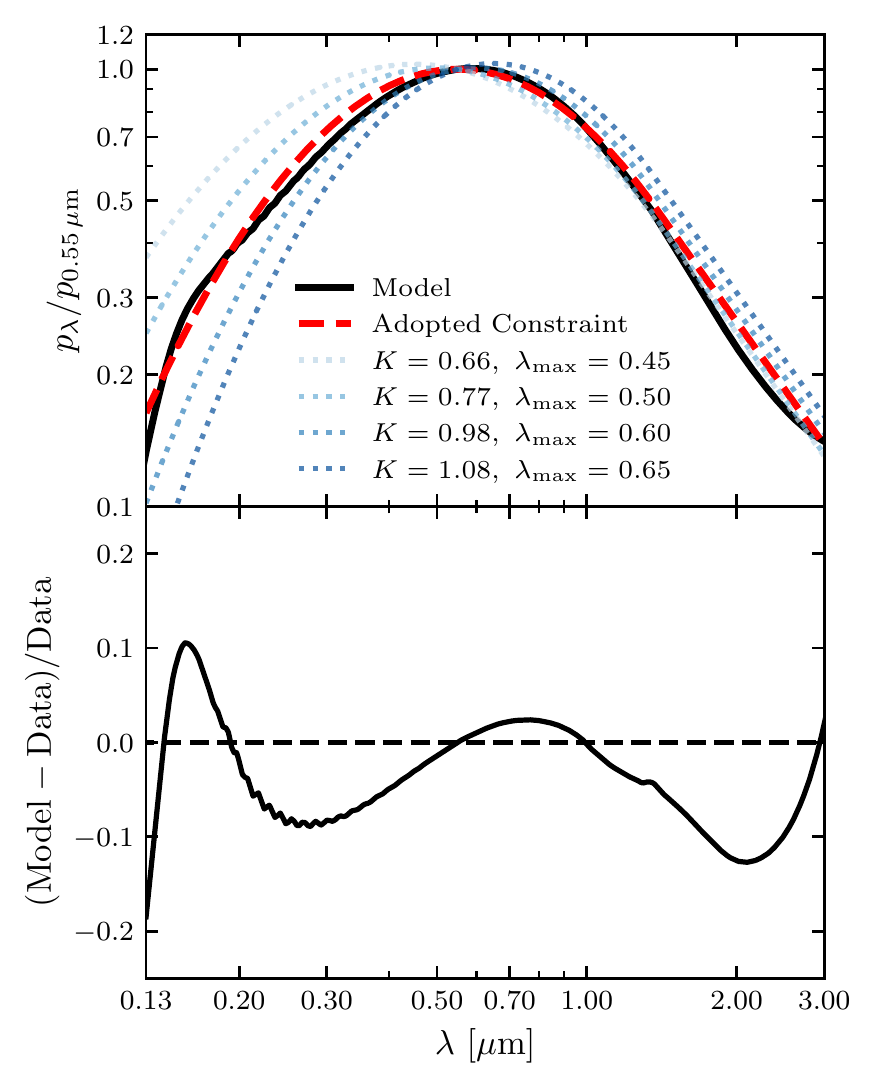}
    \caption{In the top panel, the polarized extinction profile ($p_\lambda/p_{0.55\,\mu{\rm m}}$) of the model (black solid) is compared to the adopted profile (red dashed, corresponding to $K = 0.87$ and $\lambda_{\rm max} = 0.55\,\mu{\rm m}$) as well as to a number of Serkowski Law fits from the literature \citep[blue dotted;][]{Whittet_2003}. All Serkowski Law parameterizations are extended into the infrared assuming $p_\lambda \propto \lambda^{-1.6}$ \citep{Martin+etal_1992}. The residuals with respect to the adopted constraint are plotted in the bottom panel, illustrating agreement within 10\% at nearly all wavelengths.} \label{fig:model_extpol} 
\end{figure}

\begin{figure}
    \centering
        \includegraphics[width=\columnwidth]{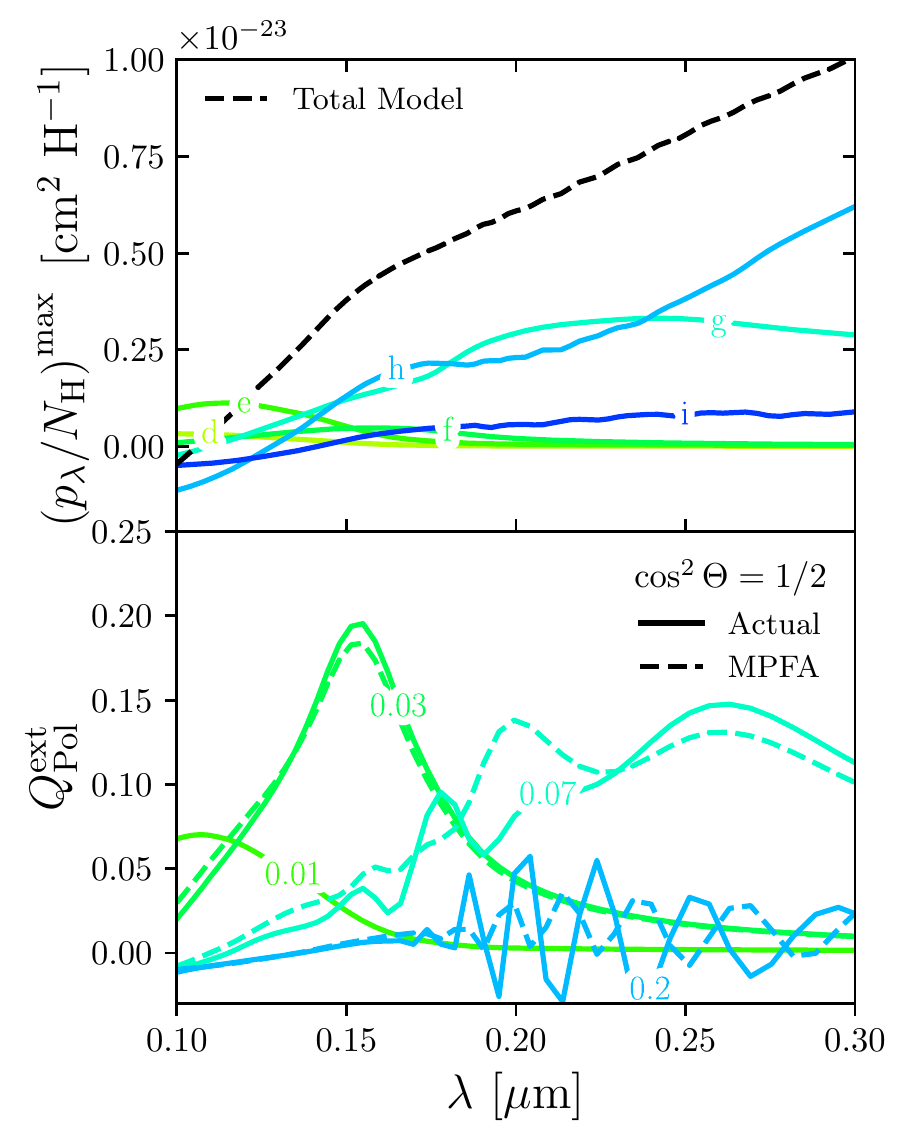}
    \caption{The top panel presents the total polarized extinction from the model (black dashed) as well as the individual contributions to the total from grains in size bins $e$--$i$ (see Figure~\ref{fig:size_analysis}). The bottom panel compares the polarized extinction efficiency computed with the MPFA (dashed lines) with the actual values (solid lines) for grains of effective radius 0.01, 0.03, 0.07, and 0.2\,$\mu$m and $\Theta = 45^\circ$. These radii fall within size bins ``e,''
    ``f,'' ``g,'' and ``h,'' respectively, in Figure~\ref{fig:size_analysis}.} \label{fig:model_extpoluv} 
\end{figure}

Figure~\ref{fig:model_summary} demonstrates that the model reproduces the principal features of interstellar polarized extinction: its amplitude relative to total extinction and to polarized emission, its broad peak at optical wavelengths, and its sharp decline toward both the UV and the IR. In agreement with previous work \citep[e.g.,][]{Kim+Martin_1995, Draine+Fraisse_2009, Siebenmorgen_2017, Guillet+etal_2018}, we find that a single dust composition suffices to account for these features. Figure~\ref{fig:model_extpol} presents the model polarization profile $p_\lambda/p_{0.55\,\mu{\rm m}}$ from the UV through the NIR, illustrating agreement with the adopted constraints to within 10\% over nearly the entire range. The discrepancies between the model and adopted constraint in the UV are in general smaller than the observed variations in the Serkowski Law.

Of the observational constraints we compare our model against in this work, polarized extinction is subject to some of the greatest systematic uncertainties given the relatively limited number of stars for which multi-wavelength polarimetry has been obtained, especially in the UV and IR. Nevertheless, there are some discrepancies of note that suggest fundamental limitations of the approach adopted here in which all aligned grains have precisely the same composition and shape.

Evident in Figure~\ref{fig:model_extpol} is a feature in the model polarized extinction at $\sim$1500\AA\ that is not present in the adopted constraints nor in observations of UV polarimetry toward individual stars \citep{Somerville+etal_1994, Martin+Clayton+Wolff_1999}. Figure~\ref{fig:model_extpoluv} illustrates the contributions of the size bins defined in Figure~\ref{fig:size_analysis} to the UV polarized extinction. Grains in size bin ``f,'' corresponding to effective radii 0.019--0.042\,$\mu$m, have a polarization efficiency $Q_{\rm pol}^{\rm ext} \equiv C_{\rm pol}^{\rm ext}/\pi a^2$ that is strongly peaked near 1500\,\AA. If grains in size bin ``f'' have nonzero alignment, as is the case for our best fit alignment function (see Figure~\ref{fig:model_dnda}), then the model can have very few grains of this size before the UV polarized extinction is over-produced. Indeed, we see in Figures~\ref{fig:model_dnda} and \ref{fig:size_analysis} that there is a pronounced minimum in the size distribution in bin ``f.''

Is this deep minimum likely to be a feature of the true interstellar grain size distribution? We expect that integrating over a distribution of grain shapes, including non-spheroidal shapes in the model, and/or considering even a modest distribution of grain dielectric functions would greatly suppress such resonant behavior in the cross sections. A further complication illustrated by Figure~\ref{fig:model_extpoluv} is the breakdown of the MPFA for certain grain sizes at these wavelengths. The MPFA is exact when $\Theta = 0$ or $90^\circ$ (see Section~\ref{subsec:cross_sections}), but not at intermediate values. For $\Theta = 45^\circ$ where the approximation is expected to be the most inaccurate, we find that the UV MPFA polarization cross sections for individual grains in size bins ``g'' and ``h'' can differ significantly from their true values, although these (oscillatory) errors are ameliorated when averaged over the continuous size distribution. Thus, we view both the 1500\,\AA\ feature in the model's polarized extinction and the deep minimum in the astrodust size distribution at $a \simeq 0.025$\,$\mu$m as potential artifacts of the simplifying assumptions of the model.

Figure~\ref{fig:model_extpoluv} demonstrates that the polarization cross sections for some grain sizes can change sign in the FUV, i.e., the polarization direction changes from parallel to the local magnetic field, to perpendicular. In our best fit model, however, a positive contribution to the polarized extinction from small grains ($a \sim0.01\,\mu$m) offsets the orthogonal polarization of the largest grains, preserving the typical polarization direction. The shortest wavelength UV polarization measurements at 1300\,\AA\ toward HD~7252 and HD~161056 show no evidence for changes in the polarization direction \citep{Somerville+etal_1994}. Such a reversal at sufficiently short wavelengths is not physically impossible, but in our view unlikely to persist after averaging over a realistic distribution of grain sizes, shapes, and angles between grain rotation axes and the local magnetic field.

\begin{figure*}
    \centering
        \includegraphics[width=\textwidth]{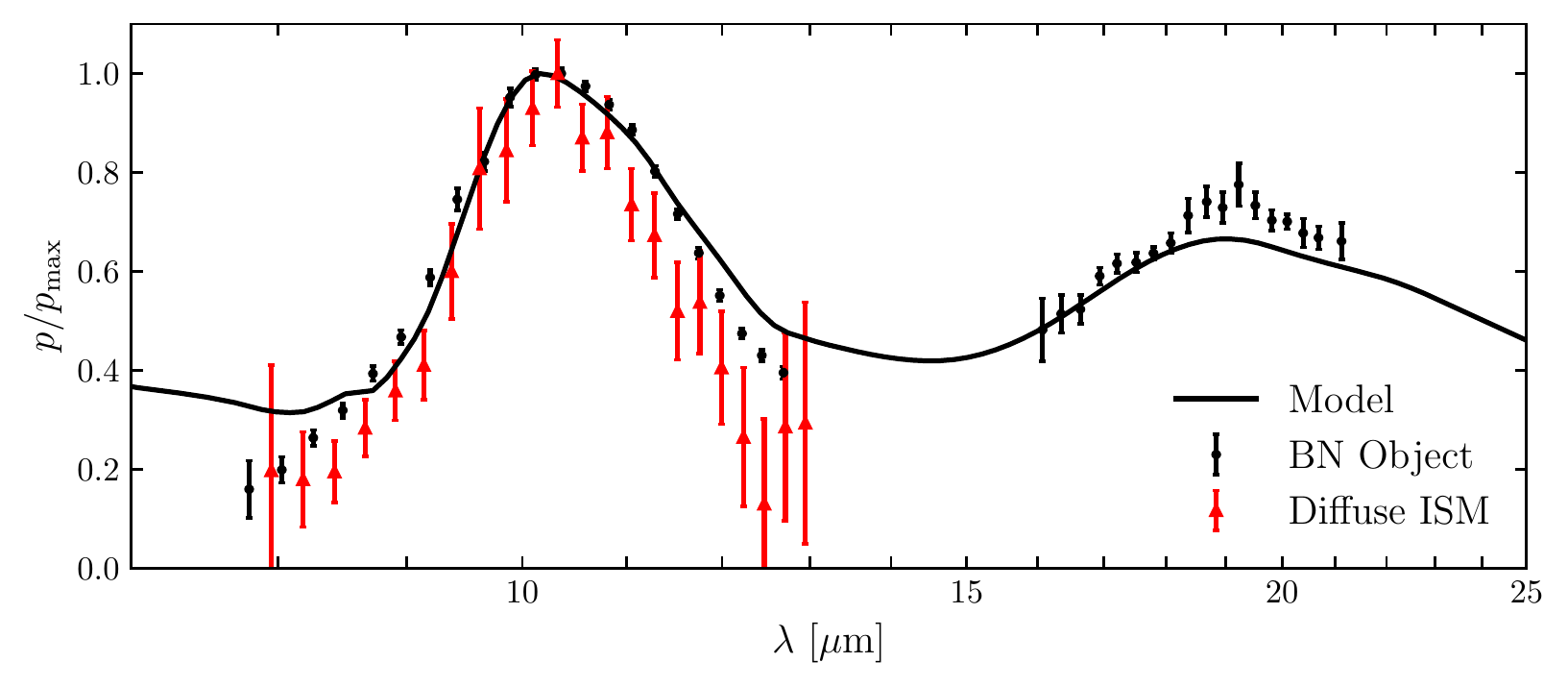}
    \caption{A comparison of the polarization profile of the silicate features in our model to observational data representative of the diffuse ISM \citep[red triangles,][]{Wright+etal_2002} and toward the BN~Object \citep[black circles,][]{Aitken+Smith+Roche_1989}.} \label{fig:model_extpolir} 
\end{figure*}

Observations of the infrared polarized extinction suggest power law behavior between $\sim1$--5\,$\mu$m with indications of flattening between 4--5\,$\mu$m \citep{Martin+etal_1992}. While generally consistent with this behavior, our best fit model has a somewhat steeper slope than the typical $p_\lambda \propto \lambda^{-1.6}$ between 1--2\,$\mu$m, and somewhat flatter between 3--4\,$\mu$m. A similar mismatch is evident in the total extinction (see Figure~\ref{fig:model_extir}). Given that both total and polarized extinction from 2--5\,$\mu$m comes mostly from grains in size bin ``i'' (0.21--0.47\,$\mu$m, see Figure~\ref{fig:size_analysis}), we anticipate that agreement could be improved by refinement of the astrodust dielectric function in this wavelength range rather than invocation of a more complicated size distribution. Additional observations of infrared polarized extinction, particularly from 3--8\,$\mu$m, would be of great value in constraining the optical properties of these grains.

The astrodust dielectric function includes an absorption feature at 3.4\,$\mu$m associated with aliphatic hydrocarbons \citep{Draine+Hensley_2021a}, and so this feature appears in model predictions for polarized extinction (see Figure~\ref{fig:model_summary}).  This feature is associated with CH bonds that may be destroyed by UV and cosmic rays, but regenerated by exposure to H atoms \citep{Mennella+etal_2003}. On this basis, \citet{Draine+Hensley_2021a} argue that the 3.4\,$\mu$m feature may be associated with grain surfaces only. Because the interstellar grain size distribution has the majority of its surface area in small unaligned grains but most of its volume in large aligned grains (see Table~\ref{table:quantities}), a feature associated with grain surfaces would be much less polarized than one arising from the full grain volume. In the formalism laid out in Section~\ref{sec:modeling}, we have not accounted for this possibility and thus the strength of the 3.4\,$\mu$m feature in the results presented here should be considered an upper limit---indeed, the strength of the feature relative to the polarized extinction at 9.7\,$\mu$m exceeds upper limits placed on the sightline toward the Galactic Center \citep{Chiar+etal_2006}. However, our model could be modified to comply with the upper limit if the material responsible for the feature is assumed to be confined to grain surface layers of thickness $\lesssim 100$\,\AA\ \citep[see discussion in][]{Draine+Hensley_2021a}.

In Figure~\ref{fig:model_extpolir}, we present the polarized extinction from the best fit model from 7--25\,$\mu$m, normalizing to unity at the maximum extinction in the 9.7\,$\mu$m silicate feature. We compare to a representative polarization profile of the 9.7\,$\mu$m feature in the diffuse ISM \citep{Wright+etal_2002} and to observations of the BN Object in both the 9.7 and 18\,$\mu$m features \citep{Aitken+Smith+Roche_1989}. Agreement in the 9.7\,$\mu$m feature is generally good. Agreement with measurements of the 18\,$\mu$m feature toward the BN Object suggests the relative strengths of the features are being modeled successfully, though with the caveat that the sightline toward the BN Object is far from diffuse and so may not be representative.

In detail, however, it is clear that the model 9.7\,$\mu$m feature profile is broader than either set of observational data plotted in Figure~\ref{fig:model_extpolir}. At issue may be an overprediction of the extinction at 8\,$\mu$m, as noted in Section~\ref{subsec:ext}, which in turn produces too much polarized extinction on the blue side of the feature. It is also the case that the observations are from a total of only three lines of sight, and so additional data on more lines of sight would be beneficial for establishing profiles representative of the diffuse ISM.

Figure~\ref{fig:model_extpolir} shows the polarization normalized to the $10$\,$\mu$m peak. With a single material (``astrodust'') providing polarization at all wavelengths, we find the ratio $p(10\,\mu{\rm m})/p(0.55\,\mu{\rm m}) = 0.22\pm0.03$, depending on the adopted axial ratio and porosity \citep{Draine+Hensley_2021c}. The present model has $p(10\,\mu{\rm m})/p(0.55\,\mu{\rm m})=0.23$. Following submission of the present paper, \citet{Telesco+etal_2022} measured the $10$\,$\mu$m polarization toward Cyg~OB2-12, finding $p(10\,\mu{\rm m})=1.24\pm0.28\%$. As this line of sight has $p(0.55\,\mu{\rm m})=9.0\pm0.3\%$ \citep{McMillan+Tapia_1977}, this yields $p(10\,\mu{\rm m})/p(0.55\,\mu{\rm m}) \approx0.14\pm0.03$, only 60\% of our predicted value---a $\sim$3$\sigma$ discrepancy with the present model.

The \citet{Telesco+etal_2022} result for $p(10\,\mu{\rm m})/p(0.55\,\mu{\rm m})$ calls into question the present model. However, Cyg~OB2-12 is a blue hypergiant with a substantial stellar wind and evidence of variability \citep[e.g.,][]{Clark+etal_2012}. It is possible that the starlight may have some intrinsic polarization at optical wavelengths, which would compromise determination of the interstellar contribution. \citet{Whittet_2015} shows that the polarization angle of the Cyg~OB2-12 sightline varies systematically with wavelength, as would be the case if the intrinsic stellar polarization differed in orientation from the interstellar polarization. If there is intrinsic stellar polarization, it may be time-dependent; additional measurements of the polarization of Cyg~OB2-12 at optical wavelengths would be valuable. Additionally, as this is the only sightline where optical and 10$\micron$ polarization have been simultaneously measured, it would be of great value to measure $p(10\,\mu{\rm m})/p(0.55\,\mu{\rm m})$ for other sightlines.

The present model has been based on what was thought to be the best estimate for the observed interstellar extinction. If the actual 5--15$\micron$ extinction is significantly lower, as found by \citet{Gordon_2021}, then the astrodust dielectric function for $\lambda > 5\,\mu$m will be appreciably modified, and the predicted $p(10\,\mu{\rm m})/p(0.55\,\mu{\rm m})$ will be lowered. Whether this change will bring the model into agreement with \citet{Telesco+etal_2022} remains to be seen; this will be the subject of a future paper.

The present work has been limited to spheroidal shapes. The physical optics differs greatly between the optical (wavelengths comparable to grain sizes, so that retardation effects matter) and the infrared (wavelengths large compared to the grain size, so that the electric field in the grain is approximated by the electrostatic solution). Hence one might anticipate that the predicted $p(10\,\mu{\rm m})/p(0.55\,\mu{\rm m})$ might change for nonspheroidal grain shapes.  Because of the difficulty of calculations for nonspheroidal grain shapes, this will require a separate dedicated study.

\subsection{Emission}
\label{subsec:irem}

\begin{figure}
    \centering
        \includegraphics[width=\columnwidth]{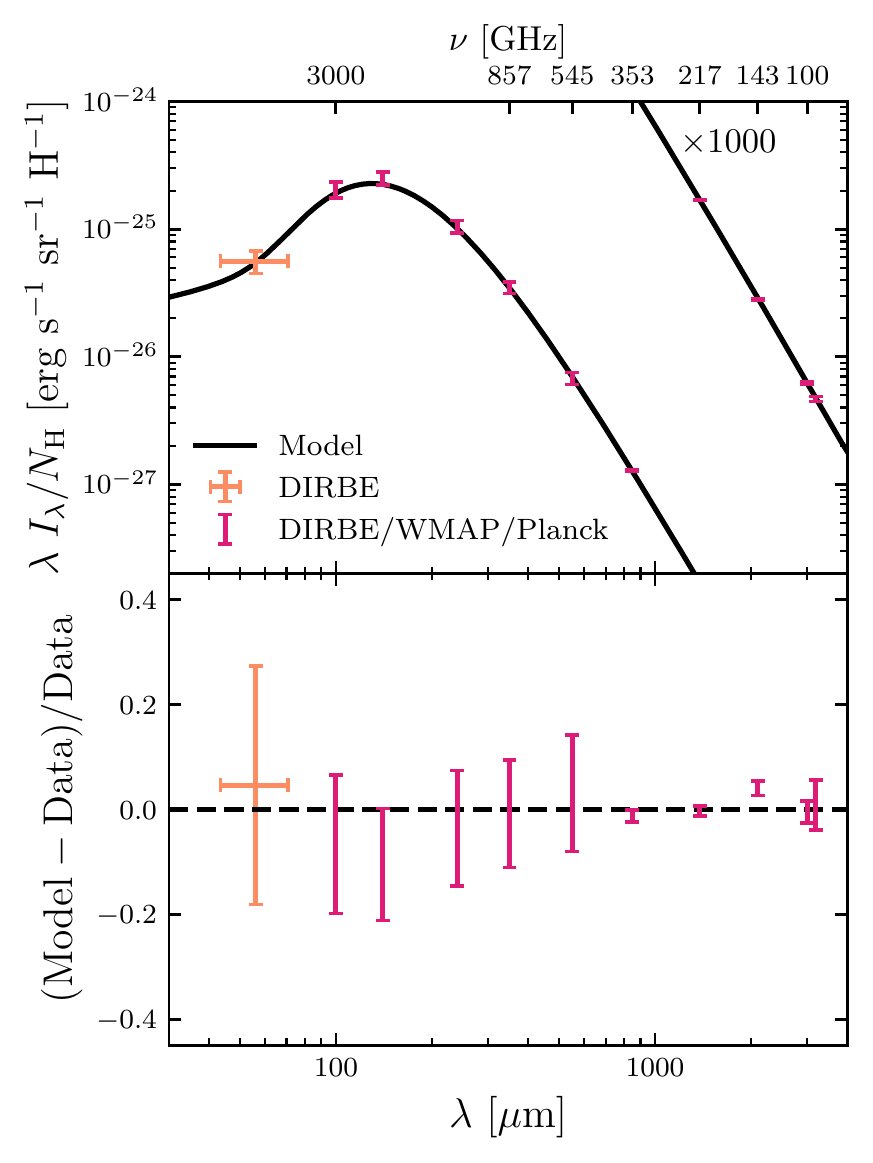}
    \caption{The top panel compares our model emission (black solid) to observations of \hi-correlated dust emission from DIRBE \citep{Dwek+etal_1997} and from analyses of DIRBE, WMAP, and Planck data \citep{Planck_Int_XVII, Planck_Int_XXII, Planck_2018_XI} as synthesized by \citet{Hensley+Draine_2021}. For clarity, the longest wavelength dust emission has been multiplied by a factor of 1000. The bottom panel quantifies deviation of the model from the data, illustrating excellent agreement.} \label{fig:model_irem} 
\end{figure}

\begin{figure}
    \centering
        \includegraphics[width=\columnwidth]{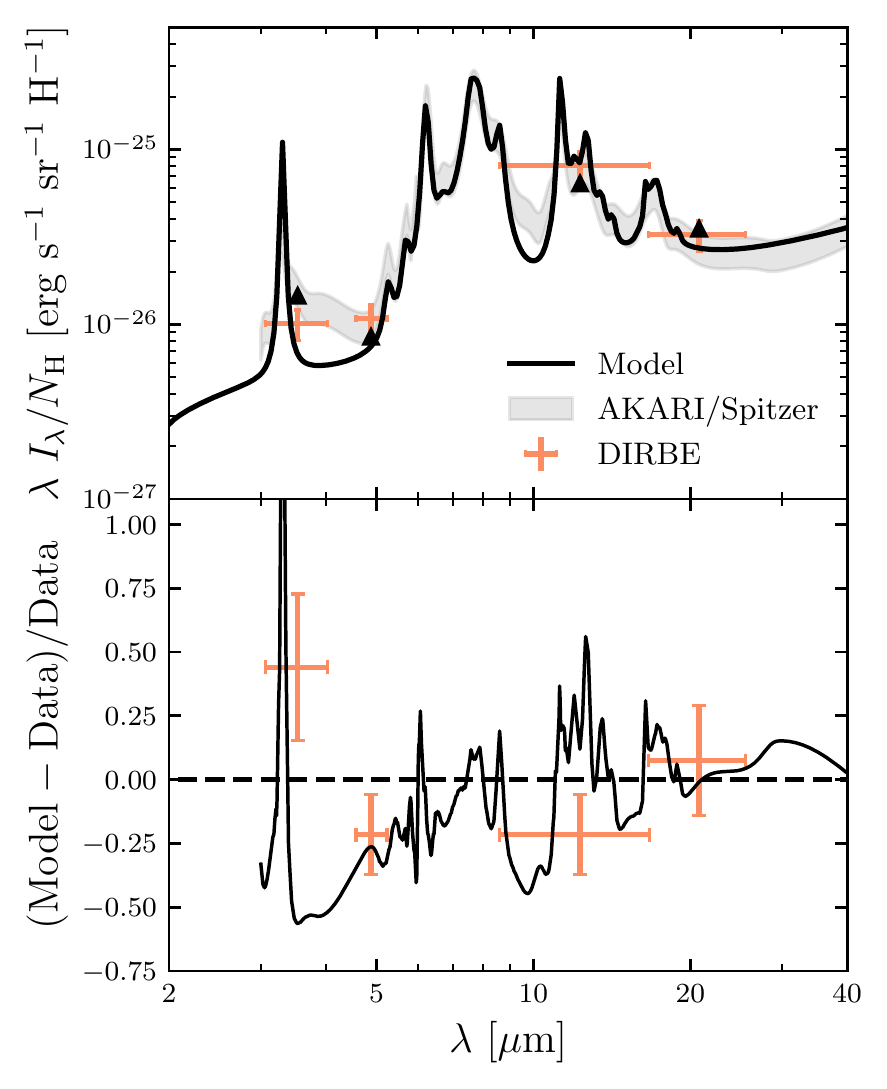}
    \caption{The top panel compares our model dust emission (black solid) to our adopted constraints based on Spitzer and AKARI data \citep[gray shaded region;][]{Ingalls+etal_2011, Lai_2020, Hensley+Draine_2021}. Also presented is the \hi-correlated emission in the DIRBE bands (orange error bars), which are compared against bandpass-integrated model values (black triangles). The lower panel quantifies the residuals.} \label{fig:model_pah} 
\end{figure}

\begin{figure}
    \centering
        \includegraphics[width=\columnwidth]{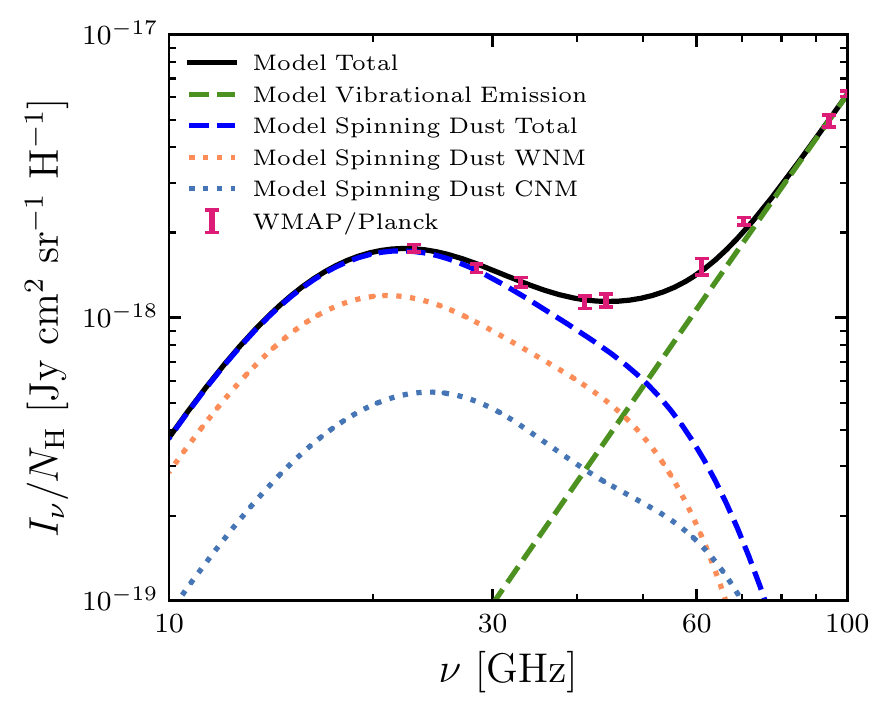}
    \caption{Comparison of observations of microwave dust emission by WMAP and Planck \citep[purple error bars,][]{Planck_Int_XVII,Planck_2018_XII} to the model (black solid line). The contributions to the model from vibrational and by rotational emission are plotted with green and blue dashed lines, respectively. The rotational emission is further broken down into contributions from the WNM (orange dotted) and CNM (blue dotted), where we have adopted a CNM fraction of 28\%.} \label{fig:model_ame} 
\end{figure}

In Figure~\ref{fig:model_irem} we compare our best fit model to the synthesized SED of \citet{Hensley+Draine_2021} based on analyses with IRAS, WMAP, and Planck data \citep{Planck_Int_XVII, Planck_Int_XXII, Planck_2018_XI}. We additionally compare to the \hi-correlated dust emission measured in the DIRBE 60\,$\mu$m band \citep{Dwek+etal_1997} by convolving the model with the DIRBE spectral response function. At all wavelengths the model provides a good fit to the data, with deviations of less than 10\%. 

Agreement at long wavelengths is not unexpected given that the astrodust dielectric function was designed to yield the observed level of FIR emission \citep{Draine+Hensley_2021a}. However, excellent agreement persists even at shorter wavelengths where there is a substantial contribution from stochastically heated grains, including both relatively small astrodust grains and PAHs. This result demonstrates that separate silicate and carbonaceous populations with different dust temperatures are not needed to account for either the width or the amplitude of the FIR dust SED. A single astrodust component suffices.

In Figure~\ref{fig:model_pah} we compare the best fit model to our adopted constraints on mid-infrared emission as well as to \hi-correlated emission measured in the DIRBE bands \citep{Dwek+etal_1997}. We find overall good agreement, although the emission from 3.5--5\,$\mu$m is underproduced by the model by roughly a factor of two. The model reproduces the observed emission in the broad DIRBE bands at a level compatible with the nominal $\sim$20\% uncertainties.

Emission in the 3.3\,$\mu$m feature is over-produced by the model relative to the adopted constraints. However, there is considerable uncertainty in strength of this feature relative to the longer wavelength PAH features in the diffuse, high latitude gas we aim to model in the present study. The observational constraints at 3.3\,$\mu$m are based on a composite spectrum of many nearby star-forming galaxies \citep{Lai_2020} that has a spectral shape compatible with observations of a Galactic translucent cloud \citep{Ingalls+etal_2011} in the range 5.3--12\,$\mu$m \citep{Hensley+Draine_2021}, but that could differ systematically in the 3.3\,$\mu$m feature where measurements of the translucent cloud are unavailable. JWST will soon clarify the relative strength of the 3.3\,$\mu$m feature in the diffuse ISM and guide refinement of the model.

Relative to the adopted constraints, the model under-produces emission in the 3.5--5.0\,$\mu$m and 9--10\,$\mu$m continuum by up to a factor of two (see Figure~\ref{fig:model_pah}). The current prescription for PAH emission includes a small amount of emission calculated with graphite cross sections to yield a modest continuum between the PAH features (see Equation~\eqref{eq:pah_graphite}). Since this is insufficient to reproduce the adopted constraints, the global optimization includes a small population of astrodust nanoparticles to help address the shortfall, as can be seen from the size bin ``a'' and ``b'' contributions to the upper left panel of Figure~\ref{fig:size_analysis}. Because we have fixed the emission properties of spinning PAHs to reproduce the AME, as we discuss below, the nanoparticle astrodust population is effectively capped and so makes only a marginal improvement to the MIR fit. In reality, we expect a heterogeneous variety of nanoparticles to contribute to the $\sim3$--20\,$\mu$m continuum, but a detailed treatment of such a population is beyond the scope of the present study.

Figure~\ref{fig:model_ame} presents the emission from the best fit model at microwave frequencies, illustrating both the total emission and the spinning dust contribution alone. The correspondence with WMAP and Planck observations is excellent. However, there are a few subtleties of the model worth exploring in detail.

While it remains the case that physical dust models have little trouble producing enough microwave emission to explain the AME, observational results over the last decade have posed challenges to the spinning PAH models put forward by \citet{Draine+Lazarian_1998a}. First, the AME spectrum of the diffuse ISM is not well fit by a spinning dust spectrum of a single idealized interstellar environment---rather, parametric fits have often resorted to a superposition of two spinning dust spectra \citep[e.g.,][]{Planck_Early_XX, Planck_2015_X}. While some authors have identified such spectra with CNM and WNM components \citep{Ysard+MivilleDeschenes+Verstraete_2010, Hoang+Lazarian+Draine_2011}, a lack of correlation between the CNM fraction and the AME frequency spectrum has challenged this interpretation \citep{Hensley+Murray+Dodici_2022}. 

Second, the spinning dust models presented by \citet{Draine+Lazarian_1998b} have a typical 30\,GHz emissivity of $\sim 10^{-17}$\,Jy\,cm$^2$\,sr$^{-1}$\,H$^{-1}$, consistent with observational data at the time \citep{Kogut+etal_1996, deOliveiraCosta+etal_1997, Leitch+etal_1997}. Subsequent analyses employing WMAP and Planck data \citep{Planck_Int_XVII, Planck_Int_XXIII}, including a recent determination that also utilizes new maps from the C-Band All-Sky Survey \citep{Harper_2022}, suggest an emissivity an order of magnitude lower. Changes to the baseline models are therefore required.

Finally, the association between the AME and PAHs has been questioned on the basis of a lack of correlation between the AME and MIR PAH emission after removing their mutual correlation with the FIR dust continuum \citep{Hensley+Draine+Meisner_2016}. In response, some authors have suggested that the AME could arise from spinning nanosilicates instead \citep{Hoang+Vinh+QuynhLan_2016, Hensley+Draine_2017b}, though \citet{Ysard_2022} have recently argued that emission from nanosilicates cannot reproduce the observed AME SED shape in detail. Another possibility advanced by \citet{Bell_2019} is that comparing broadband MIR and microwave fluxes is not as reliable as inferring parameters like PAH mass from SED fitting, and that doing so can obscure true correlations. A correlation between MIR and microwave emission could also be obscured if MIR PAH emission and spinning dust emission have markedly different emissivities in different interstellar environments, even if both are produced by PAHs \citep{Hensley+Murray+Dodici_2022, Ysard_2022}.

We find that these potential issues with the spinning PAH model can be addressed with a suitable electric dipole moment distribution. The total spinning dust emission from physical models can be readily reduced if a sizable fraction of PAHs are symmetric and have no dipole moment. While the $\sim20\,$GHz AME peak can be easily accounted for by a modest population of PAHs in the WNM with an electric dipole moment parameter $\beta_{\rm ed} = 0.3$\,D, emission at higher frequencies is sensitive to the abundance of grains with relatively low dipole moments. Our best fit model assigns $\beta_{\rm ed} = 0.3$\,D to 4\% of the PAHs, $\beta_{\rm ed} = 0.04$\,D to 35\% of the PAHs, and $\beta_{\rm ed} = 0$ to the remaining 61\%. While we do not claim this to be a robust determination of the true interstellar electric dipole moment distribution, it illustrates that there exists ample viable model space to account for observations. Given that the spinning dust emission is sensitive to the local interstellar environment in ways that PAH emission is not, we still view the spinning PAH hypothesis as the most likely explanation for the AME despite lack of evidence of correlation with the MIR emission features. Clearly, however, more theoretical and observational work is needed to clarify the nature of the AME carrier(s).

\subsection{Polarized Emission}
\label{subsec:ipol}

\begin{figure}
    \centering
        \includegraphics[width=\columnwidth]{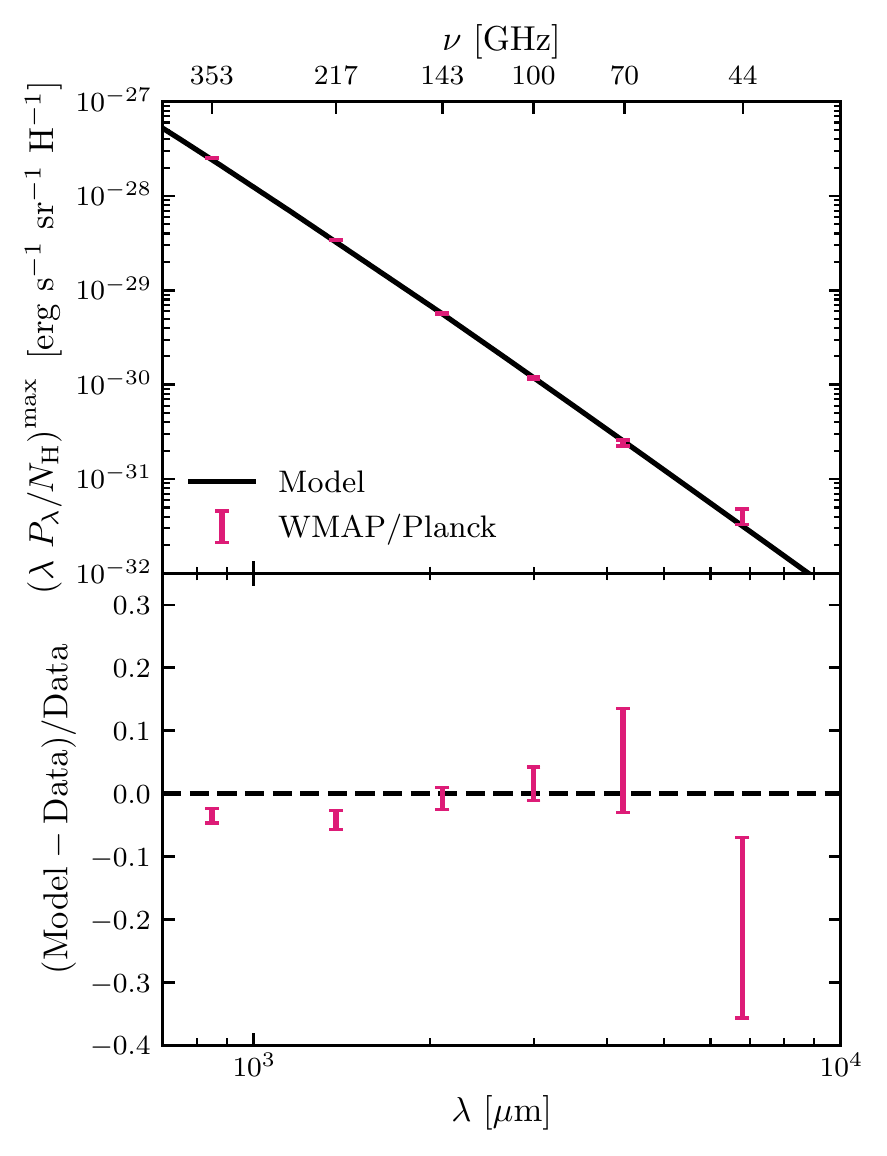}
    \caption{The top panel compares our model polarized emission (black solid) to WMAP and Planck measurements of polarized dust emission \citep[magenta error bars,][]{Planck_2018_XI}. The polarization spectrum has been scaled to its maximum value per $N_{\rm H}$ following \citet{Hensley+Draine_2021}. The bottom panel quantifies deviations of the model from the data, illustrating excellent agreement at all wavelengths.} \label{fig:model_pol}
\end{figure}

\begin{figure}
    \centering
        \includegraphics[width=\columnwidth]{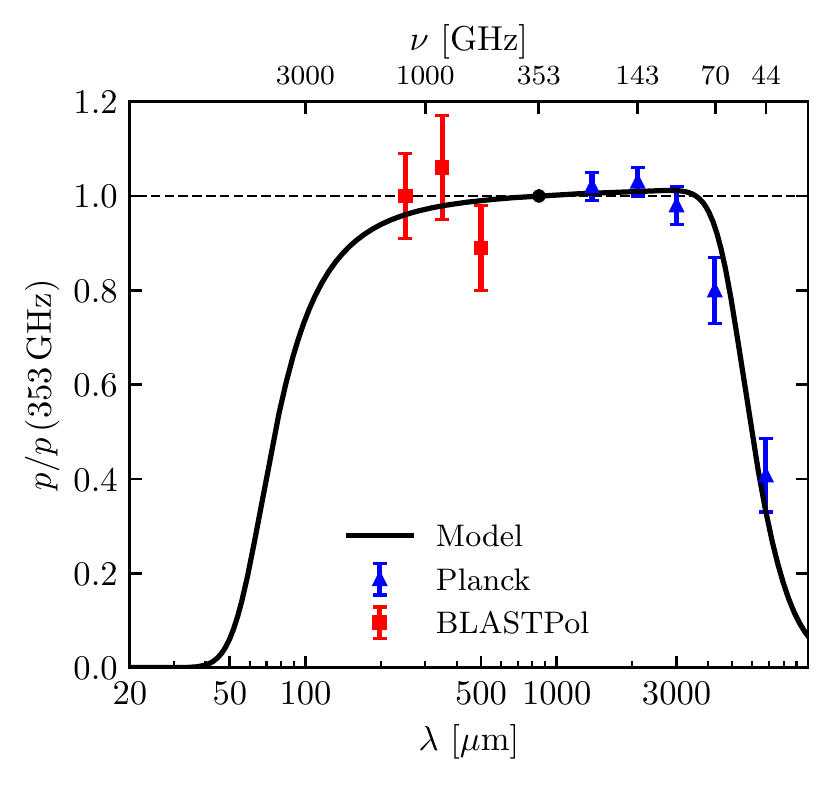}
   \caption{The frequency dependence of the fractional polarization of the dust emission is assessed by normalizing to unity at 353\,GHz. We compare observations over much of the high latitude sky by the Planck satellite \citep{Planck_2018_XI} and in the Vela~C Molecular Ridge from the BLASTPol balloon \citep{Ashton+etal_2018} to the best fit model. In the model, the sharp decline toward long wavelengths is explained by the contribution from unpolarized spinning dust emission while the decline toward short wavelengths arises from the contribution of weakly polarized emission from small, poorly aligned, stochastically heated grains.} \label{fig:model_pfrac}
\end{figure}

The polarized emission from the best fit model is compared to the adopted constraints from Planck and WMAP data \citep{Planck_2018_XI, Hensley+Draine_2021} in Figure~\ref{fig:model_pol}, illustrating good agreement over the full frequency range over which polarized emission from the diffuse ISM has been observed (44--353\,GHz). Figure~\ref{fig:model_pfrac} displays observational constraints on the dust polarization fraction relative to 353\,GHz as determined over much of the high latitude sky with Planck \citep{Planck_2018_XI} and in the Vela~C Molecular Ridge with BLASTPol \citep{Ashton+etal_2018} and also demonstrates good correspondence between the model and observations.

The model lies only a few percent below the observed emission at 353\,GHz (see Figure~\ref{fig:model_pol}) and has a 353\,GHz polarization fraction of 19.2\%. This is slightly below the value inferred from the adopted constraints of 19.6\% \citep{Hensley+Draine_2021} and from full-sky observations of 22$^{+3.5}_{-1.4}$\% \citep{Planck_2018_XII}. The polarization fraction cannot be increased by extending efficient alignment of astrodust grains to smaller sizes, because this would result in unacceptable changes to the amplitude and frequency dependence of the polarized extinction. On the other hand, while stellar polarization in a region of the sky with among the highest values of the 353\,GHz polarization fraction showed no evidence for significant departures from a typical Serkowski law \citep{Panopoulou+etal_2019}, the relationship between the frequency dependence of the optical polarized extinction and the submillimeter polarized emission is still largely unexplored. Considering uncertainties both in deriving consistent polarized extinction and emission laws and in calculating polarization from interstellar grains at wavelengths comparable to their size, we consider the level of agreement between optical and FIR polarization in the model to be satisfactory even though the 353\,GHz polarized intensity of the model falls 3$\sigma$ (but less than 4\%) below the observed value.

It is noteworthy that the polarized intensity of the model is 1.2$\sigma$ lower than the observed polarized emission at 44\,GHz (see Figure~\ref{fig:model_pol}). If any of the astrodust material were ferromagnetic, magnetic dipole emission at microwave frequencies would be polarized orthogonal to the electric dipole emission from the grain \citep{Draine+Hensley_2013}. The total polarized emission would then be reduced relative to the expectation from typical electric dipole emission at wavelengths where magnetic dipole emission is expected \citep[$\lesssim 100\,$GHz,][]{Draine+Hensley_2013}. Since our model with no ferromagnetic inclusions already produces slightly less polarized emission than is observed at low frequencies, this argues against magnetic dipole emission being a significant contribution to the total or polarized emission at these frequencies. The solid phase iron not in silicates may therefore be in the form of non-magnetic iron oxides as part of the ``astrodust'' material and/or a separate population of non-magnetic Fe-bearing nanoparticles. Alternatively, the formalism based on the phenomenological Gilbert Equation used by \citet{Draine+Hensley_2013} to predict microwave magnetic dipole emission could overestimate the true microwave emissivity of magnetic materials, allowing a substantial reservoir of metallic Fe to exist in dust without violating these constraints on polarized emission.

Although the observed polarized emission from interstellar dust appears consistent with an opacity well-described by a power law in frequency down to at least 44\,GHz \citep{Planck_2018_XI}, the polarization fraction of the emission drops precipitously toward these low frequencies (see Figure~\ref{fig:model_pfrac}). The best fit model is in excellent agreement with these results, notably the marked lack of frequency dependence of the polarization fraction between $\sim 250\,\mu$m and 3\,mm. Indeed, over this wavelength range the model monotonically increases only slightly from 0.96 to 1.01 relative to $p\left(850\,\mu{\rm m}\right)$. At longer wavelengths, spinning dust emission becomes the dominant emission mechanism. As spinning dust emission is assumed to be unpolarized in our model \citep{Draine+Hensley_2016}, it drives the 44\,GHz polarization fraction down to 33\% of its value at 850\,$\mu$m, in agreement with observations. At wavelengths shorter than $\sim 150\,\mu$m, emission from smaller, less well aligned, transiently heated grains becomes a significant fraction of the total emission (see Figure~\ref{fig:size_analysis}). The model predicts a steep drop toward short wavelengths, reaching half of the 850\,$\mu$m polarization fraction at 76\,$\mu$m and 1\% at 42\,$\mu$m.

\subsection{Scattering}
\label{subsec:scatt}

\begin{figure}
    \centering
        \includegraphics[width=\columnwidth]{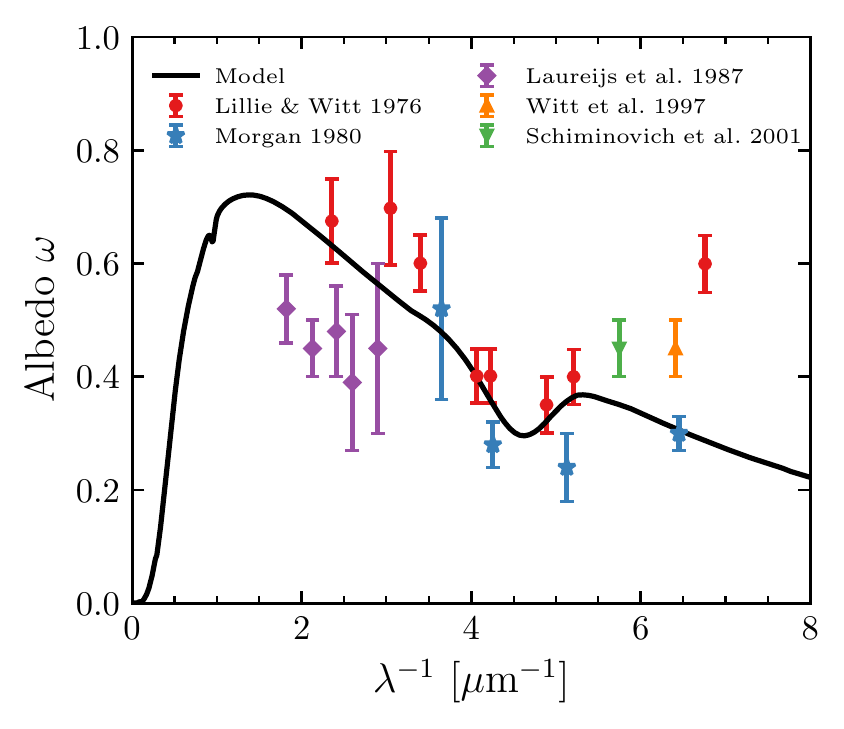}
    \caption{Comparison of the albedo of the best fit dust model (black solid line) to constraints on the albedo from observations of the diffuse Galactic light from \citet{Lillie+Witt_1976} (red circles), \citet{Witt+Friedmann+Sasseen_1997} (orange upward triangle), and \citet{Schiminovich+etal_2001} (green downward triangle). Following \citet{Guillet+etal_2018}, we supplement these with determinations of the dust albedo towards $\lambda$~Orionis \citep[][blue stars]{Morgan_1980} and Lynds~1642 \citep[][purple diamonds]{Laureijs_1987}.} \label{fig:scatt} 
\end{figure}

Observations of the diffuse Galactic light have furnished constraints on the albedo $\omega$ of interstellar dust at optical and UV wavelengths \citep{Henyey+Greenstein_1941, Lillie+Witt_1976, Witt+Friedmann+Sasseen_1997, Schiminovich+etal_2001}. Although we have not employed these data as constraints in our model fitting, we can calculate the albedo predicted by the best fit model:

\begin{equation}
    \omega = \frac{1}{\tau_\lambda/N_{\rm H}}\sum\limits_{i = 1}^N \int da \left(\frac{1}{n_{\rm H}}\frac{dn_i}{da}\right) C^{\rm sca}_{\rm ran}\left(\lambda, i,a\right)
  ~~~,
\end{equation}
where $C^{\rm sca}_{\rm ran}$ denotes the scattering cross section for randomly oriented grains and $\tau_\lambda/N_{\rm H}$ is given by Equation~\eqref{eq:ext}.

Figure~\ref{fig:scatt} demonstrates that our model is in reasonable agreement with the interstellar dust albedo inferred from observations of the diffuse Galactic light between $\sim2$--5.5$\mu$m$^{-1}$, although consistently lower than the median observed values. At wavenumbers between 5.5 and 7\,$\mu$m$^{-1}$, the model has an albedo decreasing from 0.4 to 0.3, whereas the studies of \citet{Lillie+Witt_1976}, \citet{Witt+Friedmann+Sasseen_1997}, and \citet{Schiminovich+etal_2001} indicate that the albedo rises to values of 0.4--0.6. However, \citet{Morgan_1980} found an albedo at 6.5\,$\mu$m$^{-1}$ for the dust near $\lambda$~Orionis that is generally consistent with our model---further observational studies to clarify the UV scattering properties of interstellar dust would be valuable. As the model also underpredicts the total extinction in the 5.5--7\,$\mu$m$^{-1}$ range (see Figure~\ref{fig:model_extopuv}), it is likely that the adopted astrodust and/or PAH optical properties or the adopted size distributions at these wavelengths are not quite correct.

\section{Discussion}
\label{sec:discussion}

\subsection{Implications for Dust Evolution}
\label{subsec:implications}
The present paper approximates the heterogeneous population of interstellar grains in the diffuse ISM as consisting of two grain types: ``astrodust'' particles that dominate the grain mass and optical-IR extinction, plus PAH-like nanoparticles accounting for $\lesssim10\%$ of the dust mass, but contributing significantly to UV extinction and certain characteristic MIR features (both in emission and absorption). Given the complex interplay of processes involved in the formation and evolution of interstellar dust, we must ask: is it reasonable to approximate most of the dust mass by the single composition and porosity assumed for the ``astrodust''?

The interstellar grain population includes a heterogeneous mixture of ``stardust'' particles. These stardust grains have a variety of compositions, as is known from observations of dusty outflows today \citep[from AGB stars, planetary nebulae, and supernovae;][]{Buchanan_2009, Groenewegen+etal_2009, Sloan_2017} and study of presolar grains trapped in meteorites at the time of formation of the Solar System \citep{Nittler+Ciesla_2016}. After injection into the ISM, stardust grains are subject to destruction by sputtering and grain-grain collisions in interstellar shock waves. 

Grain lifetimes $\tau_{\rm dest}$ against destruction in the ISM
remain uncertain, with estimates of 1--$20\times10^8$\,yr \citep{Barlow_1978a, Draine+Salpeter_1979, Dwek+Scalo_1980, Jones+etal_1994, Bocchio+etal_2014} with a recent paper \citep{Zhukovska+etal_2016} using $\tau_{\rm dest}=3.5\times10^8$\,yr. In the diffuse ISM, most of the refractory elements are observed to be depleted from the gas phase, and must be in solid dust grains, but---given the short estimates for $\tau_{\rm dest}$---we estimate that only $\sim$5--10\% of the grain mass is in not-yet-destroyed stardust grains. The remaining 90--95\% of the dust mass consists of materials that have been grown in the ISM.  

The larger ``astrodust'' particles are presumed to be the end-product of a number of competing processes  \citep[see, e.g.,][]{Draine_1990, Draine_2009}: (1) depletion of atoms from the gas phase onto grain surfaces, (2) ultraviolet photolysis of the accreted matter to create amorphous silicates and hydrocarbon materials (and perhaps other compounds), (3) coagulation of smaller grains to create more massive grains, and (4) occasional shattering of larger grains in grain-grain collisions.  

The observed strength of the interstellar 9.7\,$\mu$m ``silicate'' feature appears to require that accretion of atoms and ions, followed by UV photolysis, puts a large fraction of the accreted Si into amorphous silicate material. The observed ubiquity of PAH emission in normal star-forming galaxies suggests that the accreted C atoms, in the presence of arriving H atoms and UV radiation, must form PAH material. Subsequent fragmentation in grain-grain collisions can plausibly maintain the observed population of free-flying PAH nanoparticles.

Coagulation is expected in grain-grain collisions with relative velocities of $\lesssim0.1$\,km\,s$^{-1}$, while shattering seems likely for collision speeds of $\gtrsim1$\,km\,s$^{-1}$ \citep{Jones+etal_1996}. Coagulation at low collision speeds would be expected to produce highly porous ``fluffy'' grains, with ballistic aggregation of spherical monomers leading to porosities $\mathcal{P}\approx 0.87$ \citep{Shen+etal_2008}. However, the observed polarization properties of interstellar dust appear to limit porosities to $\mathcal{P}\lesssim 0.6$ \citep{Draine+Hensley_2021c}. Highly porous grains are fragile and may be less likely to survive grain-grain collisions. Suprathermal rotation driven by radiative torques may preferentially disrupt highly porous grains \citep{Silsbee+Draine_2016, Hoang_2019}. Some grain-grain collisions may lead to ``crushing'' and compactification of initially porous grain material that escapes fragmentation; this compactification may account for the apparently modest porosities of interstellar grains.

Fragmentation seems likely to lead to heterogeneous nanoparticle fragments, that presumably include the PAH nanoparticles. Because the nanoparticles dominate the grain surface area, most accretion presumably takes place on their surfaces \citep{Weingartner+Draine_1999}. Coagulation and compactification of the nanoparticles then creates aggregates of the different free-flying particles---which we term ``astrodust'' material. Models that seek to explain the abundance and size distribution of interstellar grains as resulting from stochastic sputtering in fast shocks, accretion from the gas, and coagulation and fragmentation in grain-grain collisions seem promising \citep[e.g.,][]{Hirashita+etal_2021}, although having many uncertain adjustable parameters. Nevertheless, this general picture seems compatible both with observations of the ISM and with the present grain model, where the larger grains (dominating the total grain mass) are approximated as a single population of particle containing both silicate and carbonaceous materials---``astrodust.''

\subsection{Comparison with Previous Models}
\label{subsec:model_compare}

The model presented in this work marks a strong departure from previous dust models. In this section we first highlight the similarities and differences of our model with respect to the model of \citet{Draine+Li_2007} (hereafter DL07), and then present a comparison with other models in the literature.

\subsubsection{Comparison with DL07}

\begin{figure*}
    \centering
        \includegraphics[width=\textwidth]{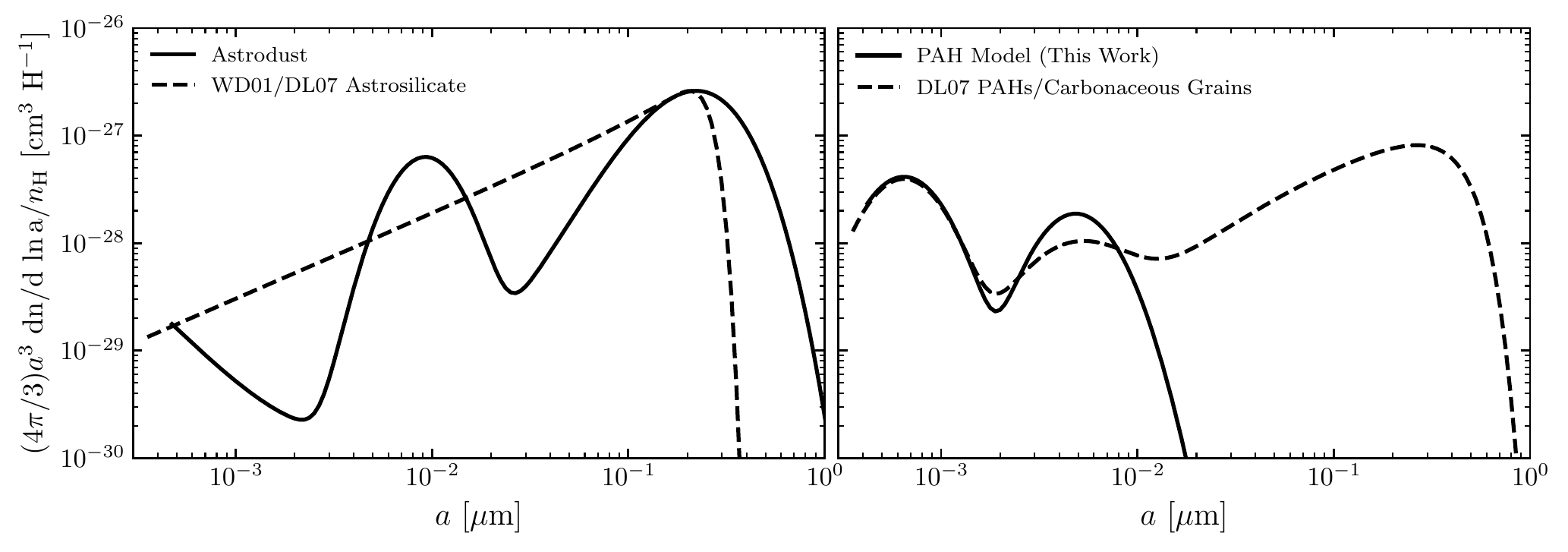}
    \caption{A comparison of the size distributions of astrodust/astrosilicte (left) and carbonaceous grains (right) between the present model (solid) and DL07 (dashed). Unlike DL07, the present model includes no large carbonaceous grains.} \label{fig:dnda_dl07} 
\end{figure*}

\begin{figure}
    \centering
        \includegraphics[width=\columnwidth]{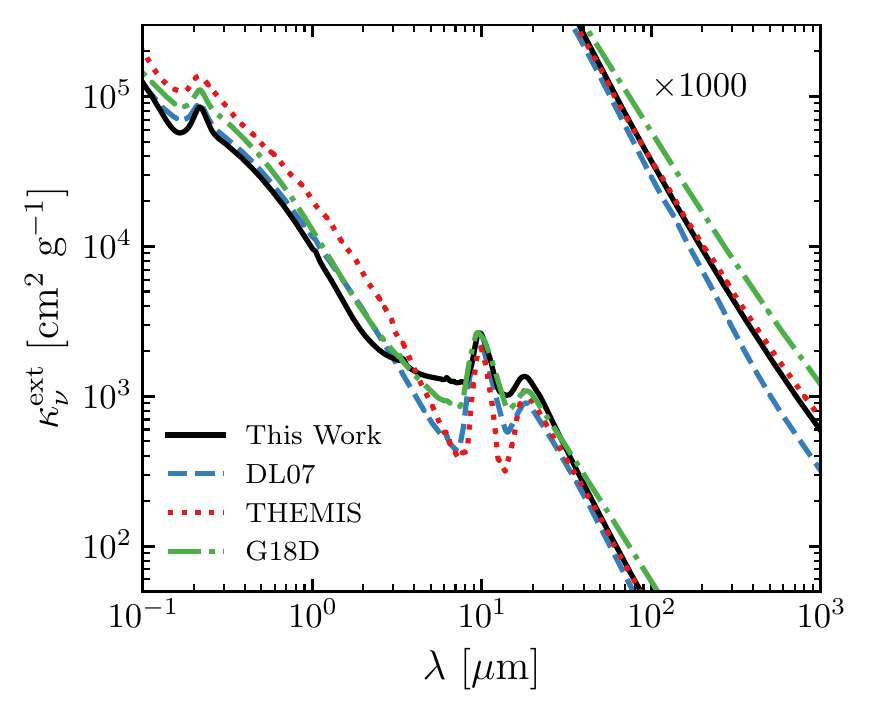}
    \caption{Total extinction per unit dust mass (see Equation~\eqref{eq:kappa}) for the astrodust model of this work (black solid), DL07 \citep[blue dashed;][]{Draine+Li_2007}, THEMIS \citep[red dotted;][]{Jones+etal_2013}, and G18 Model~D \citep[green dot-dashed;][]{Guillet+etal_2018}.} \label{fig:opacity_cf} 
\end{figure}

\begin{figure}
    \centering
        \includegraphics[width=\columnwidth]{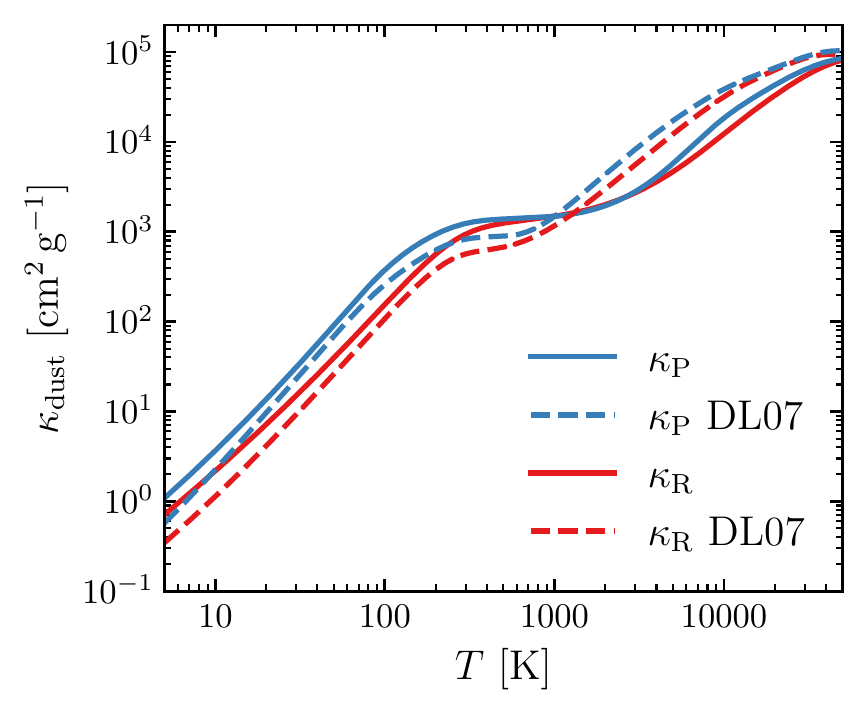}
    \caption{Planck (blue) and Rosseland (red) mean dust opacities for present model (solid) and for DL07 (dashed). Note that the plotted curves represent total dust absorption cross section per total dust mass, i.e., not including any contribution from the gas phase.} \label{fig:opacities} 
\end{figure}

\begin{figure}
    \centering
        \includegraphics[width=\columnwidth]{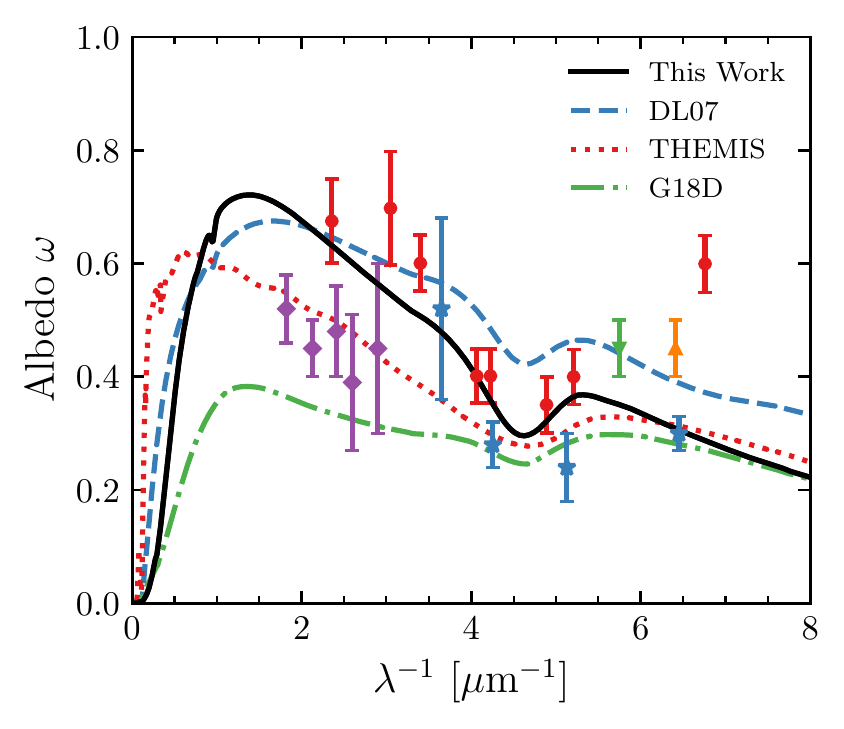}
    \caption{Comparison of the dust albedos of the astrodust model of this work (black solid), DL07 \citep[blue dashed;][]{Draine+Li_2007}, THEMIS \citep[red dotted;][]{Jones+etal_2013}, and G18D \citep[green dot-dashed;][]{Guillet+etal_2018}. The observational data from Figure~\ref{fig:scatt} are overlaid.} \label{fig:albedo_cf} 
\end{figure}

The silicate and graphite components of the DL07 ``silicate-graphite-PAH model'' originate in the model of \citet{Draine+Lee_1984}. For the silicate component, \citet{Draine+Lee_1984} derived a dielectric function consistent with the Kramers-Kronig relations that reproduced the 9.7\,$\mu$m feature profile observed in the Trapezium region \citep{Forrest+Gillett+Stein_1975}, the 18\,$\mu$m feature profile observed toward the Galactic Center \citep{McCarthy+etal_1980}, and, together with the emission from graphite grains in the model, the infrared emission in Galactic cirrus observed by IRAS \citep{Low+etal_1984}. Following the identification of the infrared emission features with polycyclic aromatic hydrocarbons \citep[PAHs,][]{Leger+Puget_1984, Allamandola+Tielens+Barker_1985}, the carbonaceous component was modified to include PAHs at small sizes ($\lesssim 50\,$\AA) and to retain graphitic properties at large sizes \citep{Li+Draine_2001b, Weingartner+Draine_2001, Draine+Li_2007}. Subsequent improvements to the astrosilicate dielectric function included extension into the X-ray \citep{Laor+Draine_1993}, smoothing of an absorption feature at 6.5\,$\mu$m$^{-1}$ not found in interstellar observations \citep{Kim+Martin_1995b, Weingartner+Draine_2001}, adjustment of the FIR opacity to better reproduce the observations of COBE-FIRAS \citep{Li+Draine_2001b}, and refinement of the X-ray absorption edges associated with C, O, Mg, Si, and Fe \citep{Draine_2003b}.

Both the ``astrodust'' of the present model and the ``astrosilicate'' of DL07 and predecessors represent materials with properties inferred from interstellar observations, hence the ``astro-'' prefix. The adopted properties are often guided by, but are not driven by, laboratory data. In synthesizing the MIR astrodust dielectric function, \citet{Draine+Hensley_2021a} used Spitzer InfraRed Spectrograph observations on the sightline toward Cyg~OB2-12 that included the 9.7 and 18\,$\mu$m silicate features \citep{Hensley+Draine_2020}. This line of sight is representative of the diffuse ISM \citep{Whittet_2015}, unlike the Trapezium or Galactic center regions used for the silicate features in the astrosilicate model. In the FIR, the astrodust dielectric function was derived to reproduce observations from the Planck satellite \citep{Planck_Int_XVII, Planck_Int_XXII}. As the 3.4\,$\mu$m absorption feature is attributed to aliphatic carbon materials incorporated in astrodust, the astrodust dielectric function was derived to reproduce Infrared Space Observatory measurements of absorption in this feature toward Cyg~OB2-12 \citep{Hensley+Draine_2020}. This feature is absent in astrosilicate. The astrodust dielectric function at optical to X-ray wavelengths is intended to represent a mixture of silicate and carbonaceous material, including the C K edge at 284\,eV \citep[see][Figures~18 and 19]{Draine+Hensley_2021a}.

The astrodust size distribution of the present model was derived from a different parameterization than employed for astrosilicate by \citet{Weingartner+Draine_2001} and \citet{Draine+Li_2007}. We compare the astrodust and DL07 astrosilicate size distributions in the left panel of Figure~\ref{fig:dnda_dl07}. Both size distributions peak at a grain size of roughly 0.2\,$\mu$m, though the astrodust size distribution does not fall off as sharply toward larger sizes. As discussed in Section~\ref{sec:summary}, the astrodust grain size distribution has much more structure at smaller grain sizes than the simple power law behavior of the astrosilicate size distribution. This is driven by the constraints on polarized extinction, as discussed in Section~\ref{subsec:extpol} and present also in astrosilicate size distributions derived from fits to polarized extinction \citep{Kim+Martin_1995, Draine+Fraisse_2009}. While the astrodust size distribution includes a small population of nanoparticles peaking at the minimum grain size, this population is still smaller in every size bin than the DL07 astrosilicate grains.

The present model employs PAH cross sections identical to those of DL07 with the exceptions of the 14.19\,$\mu$m feature and the 17\,$\mu$m complex, which have both been increased in strength by 33\% \citep[see][Appendix~A]{Draine_2021}. While the function used to interpolate between PAH and graphitic cross sections (Equation~\eqref{eq:pah_graphite}) is the same between the two models, the graphite cross sections employed here are those derived for turbostratic graphite by \citet{Draine_2016}, whereas DL07 used those of \citet{Draine+Lee_1984} with modifications at X-ray wavelengths following \citet{Laor+Draine_1993} and \citet{Draine_2003b}. Where DL07 used a non-parametric $f_{\rm ion}\left(a\right)$ based on theoretical calculations \citep{Li+Draine_2001b}, we employ Equation~\eqref{eq:fion}, which is a close approximation. DL07 assumed a mass density of 2.2\,g\,cm$^{-3}$ for all carbonaceous grains as is appropriate for bulk graphite, whereas we use 2.0\,g\,cm$^{-3}$ more representative of PAHs. This leads to a slight increase in the minimum PAH grain size from 3.5\,\AA\ in DL07 to 4.0\,\AA\ in the present model \citep{Draine_2021}. 

As in DL07, we adopt the functional form of the PAH size distribution proposed by \citet{Li+Draine_2001b}, but we employ different values for many of the parameters based on our global optimization. The size distributions for carbonaceous grains in the present model and in DL07 are compared in the right panel of Figure~\ref{fig:dnda_dl07}. Given that nearly identical PAH cross sections were adopted by both models, it is unsurprising to see close agreement in the size distribution of the smallest PAHs. The present model includes somewhat more large PAHs than DL07, driven mostly by the MIR continuum emission (see Figure~\ref{fig:size_analysis}), but includes none of the large carbonaceous grains that account for most of the carbonaceous grain mass in DL07.

Figure~\ref{fig:albedo_cf} compares the dust albedo between the present model and DL07. While the DL07 model has overall better agreement with assorted observational data longward and shortward of the 2175\,\AA\ feature, the DL07 albedo is much higher in the feature than inferred from observations of the diffuse Galactic light \citep{Lillie+Witt_1976}. In contrast, the present model fits the data in the vicinity of the 2175\,\AA\ feature well, but predicts somewhat lower albedos than observed longward and shortward of the feature (see discussion in Section~\ref{subsec:scatt}). Neither model reproduces the very high albedo of $0.60\pm0.05$ at 1480\,\AA\ reported by \citet{Lillie+Witt_1976}.

The total dust to gas mass ratio of DL07 is 0.0104, higher than the 0.0071 of the present model. The DL07 astrosilicate component has a dust to gas mass ratio of 0.0076, only slightly higher than the 0.0064 of the astrodust component of the present model. The most striking disparity is in the carbonaceous component, which has a dust to gas mass ratio of 0.0027 in DL07 but only 0.0007 in the present model, which has no large carbonaceous grains. Thus to produce the same amount of extinction per H atom as DL07, the present model requires a larger opacity.

In Figure~\ref{fig:opacity_cf}, we compare the dust opacity for total extinction $\kappa^{\rm ext}_\nu$ between the present model and DL07. Explicitly, the dust opacity at frequency $\nu$ summed over all dust components is

\begin{equation} \label{eq:kappa}
    \kappa^{\rm ext}_\nu = \frac{\tau_\nu}{N_{\rm H}}\left({\sum\limits_{i = 1}^N \rho_i V_i}\right)^{-1}
    ~~~,
\end{equation}
where $\rho_i$ is the mass density of component $i$, $V_i$ is the grain volume per H atom in component $i$ (see Equation~\eqref{eq:vol}), and $\tau_\nu/N_{\rm H}$ is given by Equation~\eqref{eq:ext}. Note that Equation~\eqref{eq:kappa} defines $\kappa_\nu$ as the extinction cross section per unit {\it dust} mass.

Overall, $\kappa_\nu$ is quite similar between DL07 and the present model, having the most significant deviations from $\sim3$--8\,$\mu$m, where the astrodust model produces much more extinction per unit mass. The DL07 model has a slightly larger opacity in the NIR, slightly smaller in the FIR, and roughly the same in the optical and UV. Consequently, we expect dust masses inferred from fitting DL07 to FIR emission to change only modestly compared to fits with the present model. Indeed a study using a preliminary version of the present model found agreement in inferred dust masses within $\sim25$\% of those inferred from fits with DL07 \citep{Chastenet_2021}. At optical wavelengths, the slightly lower opacity and dust to gas ratio of the present model implies less extinction per H atom than DL07, consistent with our adoption of a lower value of $E(B-V)/N_{\rm H}$ from \citet{Lenz+Hensley+Dore_2017} more representative of higher Galactic latitudes. Thus the present model addresses a known limitation of DL07 of predicting too much optical extinction per FIR emission \citep{Dalcanton+etal_2015, Planck_Int_XXIX}.

From $\kappa_\nu$ we can derive the Planck and Rosseland mean opacities, respectively, of the models:

\begin{align}
    \kappa_{\rm P}\left(T\right) &\equiv \frac{\pi}{\sigma_{\rm SB} T^4}\int d\nu \kappa_\nu^{\rm abs} B_\nu\left(T\right) \\
    \kappa_{\rm R}\left(T\right) &\equiv \frac{\int d\nu \frac{\partial B_\nu\left(T\right)}{\partial T}}{\int d\nu \frac{1}{\kappa_\nu^{\rm abs}} \frac{\partial B_\nu\left(T\right)}{\partial T}}
    ~~~,
\end{align}
where $\sigma_{\rm SB}$ is the Stefan-Boltzmann constant. These quantities represent frequency-averaged absorption cross sections per dust mass, i.e., not considering contributions by the gas phase.

Figure~\ref{fig:opacities} illustrates the Planck and Rosseland mean dust opacities in total extinction as a function of temperature from 5 to $5\times10^4$\,K for both the present model and DL07. Overall the models are in good agreement, with differences attributable to the greater opacity of the astrodust model from 3--8\,$\mu$m and the slightly larger opacity of DL07 at optical and NIR wavelengths (see Figure~\ref{fig:opacity_cf}). The opacities plotted in Figures~\ref{fig:opacity_cf} and \ref{fig:opacities}, as well as the frequency-dependent absorption and scattering opacities, are made available in tabular form.

\subsubsection{Comparison with Other Models}

The {\sc THEMIS} dust model \citep{Jones+etal_2017} takes another approach to dust modeling by attempting to capture the effects of the interstellar environment on the composition of dust grains (e.g., UV photo-processing, coagulation). This model employs size-dependent optical constants, drawing on laboratory data where possible \citep{Jones_2012a, Jones_2012b, Jones_2012c, Jones+etal_2013}. The principal components are aromatic carbon nanoparticles, large grains having aliphatic hydrocarbon cores and hydrogenated amorphous carbon mantles, and large grains having amorphous silicate cores, aromatic carbon mantles, and Fe and FeS nano-inclusions. While this model has been developed to reproduce the observed emission and extinction from interstellar grains, it has not yet been extended to polarization. 

All quantities for the THEMIS model presented here are based on the publicly available extinction curve\footnote{\url{https://www.ias.u-psud.fr/DUSTEM/EXT_J13.RES}} with a dust to gas mass ratio of 0.0066 \citep[see][Table~1]{Jones+etal_2017}.

\citet{Guillet+etal_2018} presented a suite of models developed to be consistent with observations of FIR dust emission and polarization from the Planck satellite. We focus here on their Model~D, hereafter G18D, which provides the best fit to the data. This model is an extension of that developed by \citet{Compiegne+etal_2011} and consists of PAHs, amorphous carbon, and astrosilicate components. The PAH model is based on that of \citet{Draine+Li_2007} with some modifications of feature cross sections, elimination of the 1.05 and 1.26\,$\mu$m bands, and addition of FIR bands following \citet{Ysard+Verstraete_2010}. The amorphous carbon grains are modeled following the optical properties of the ``BE'' sample of \citet{Zubko+etal_1996} (derived from burning benzene in air), while the astrosilicate is that used by \citet{Weingartner+Draine_2001} without the subsequent modifications employed by DL07. To achieve better agreement with the FIR emission and polarization, the G18D astrosilicate component incorporates amorphous carbon inclusions that constitute 6\% of its total volume.

We use the DustEM software\footnote{\url{https://www.ias.u-psud.fr/DUSTEM/}} \citep{Compiegne+etal_2011} on the provided G18D parameter file\footnote{GRAIN\_G17\_ModelD.DAT} to compute all relevant quantities presented here, adopting a dust to gas mass ratio of 0.0081 where needed \citep[see][Table~3]{Guillet+etal_2018}.

Figure~\ref{fig:opacity_cf} compares the frequency-dependent opacities of the THEMIS and G18D models to the astrodust model of this work and to DL07. Differences between the models can arise from the different dust to gas mass ratios (spanning 0.0066 for THEMIS to 0.0104 for DL07), adoption of different observational constraints to calibrate the models, and imperfections in the abilities of the models to reproduce those constraints. Both the G18D model and especially the THEMIS model have a significantly higher opacity at UV, optical, and NIR wavelengths than either the astrodust model or DL07. This could in part be the result of calibration to higher values of $A_V/N_{\rm H}$ from \citet{Bohlin+Savage+Drake_1978} and \citet{Rachford+etal_2009} for THEMIS \citep{Jones+etal_2013} and G18D, respectively, than to the $\sim33$\% lower value from \citet{Lenz+Hensley+Dore_2017} used in this work. While the models agree on the opacity in the 9.7\,$\mu$m feature, they diverge sharply in the 3--8\,$\mu$m range and, to a lesser extent, in the region between the 9.7 and 18\,$\mu$m features. All models are in relatively good agreement in the FIR, though the FIR opacity law of G18D is flatter than the other models, leading to factor $\sim2$ differences at $\lambda \gtrsim 500\,\mu$m.

The models differ more sharply in their predictions of the dust albedo at optical and UV wavelengths, as illustrated by Figure~\ref{fig:albedo_cf}. Both the THEMIS and G18D models fall significantly below the determinations of \citet{Lillie+Witt_1976} from the diffuse Galactic light on both the short and long wavelength sides of the 2175\,\AA\ feature. The G18D and THEMIS models are in better agreement with determinations of the dust albedo toward $\lambda$~Orionis \citep{Morgan_1980} and Lynds~1642 \citep{Laureijs_1987}. Given the power of optical and UV scattering to discriminate among these models and the level of disagreement among the observations, new determinations of the dust albedo on diffuse sightlines would be of great value.

\subsection{Theoretical and Observational Prospects}

\subsubsection{Testing the One Component Hypothesis}

\begin{figure}
    \centering
        \includegraphics[width=\columnwidth]{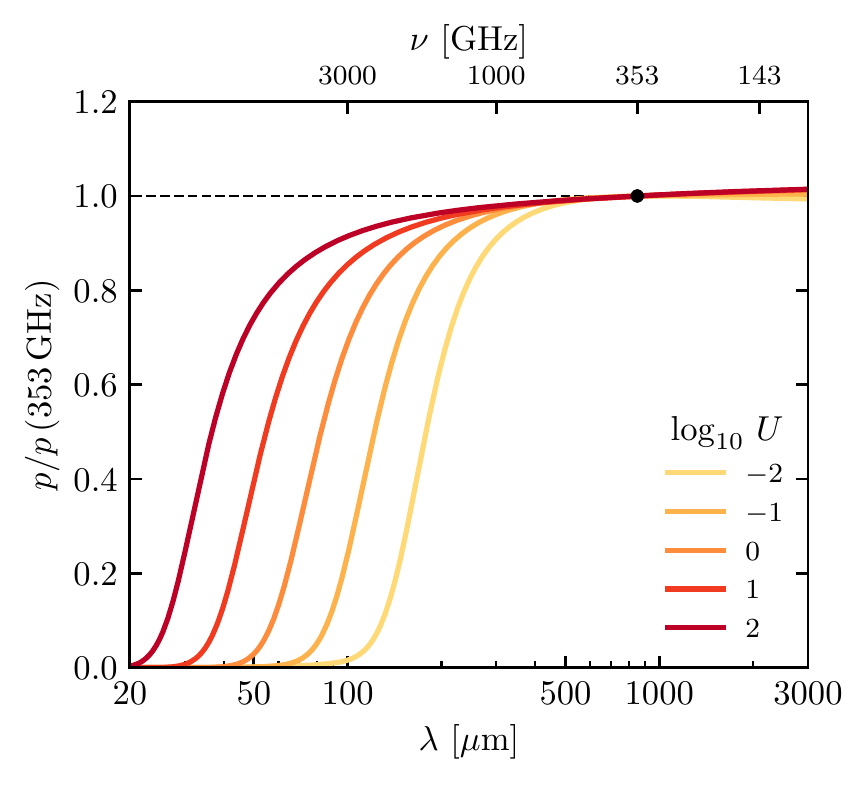}
    \caption{The polarization fraction of the dust emission in the model relative to its value at 353\,GHz is plotted for different values $U$ of the intensity of radiation heating the dust (see Equation~\eqref{eq:U}). A contribution from spinning dust emission is not included.} \label{fig:pfrac_U}
\end{figure}

The one component hypothesis put forward by this study was prompted in large part by measurements of the polarized dust emission at FIR/submm wavelengths. Little difference is observed between the dust SED in total versus polarized intensity (see Figure~\ref{fig:model_pfrac}), which is difficult to reconcile with multiple dust populations having generically different compositions, temperatures, and alignment efficiencies \citep[see discussion in][]{Draine+Hensley_2021a}. However, with the exception of a few translucent regions measured with the BLASTPol balloon experiment \citep{Ashton+etal_2018, Shariff+etal_2019}, polarization measurements of diffuse lines of sight have been limited to low frequencies ($\nu \leq 353$\,GHz) where the temperature effects that can distinguish different dust compositions are less important. Hence, an important test of the one component hypothesis is polarimetry near the peak of the dust SED ($60 \lesssim \lambda/\mu{\rm m} \lesssim 300\,\mu$m) on diffuse sightlines \citep{Hensley_2019}.

In Figure~\ref{fig:pfrac_U} we present our model predictions for the frequency dependence of the dust polarization fraction $p$ as a function of the intensity of the illuminating radiation. Illustrated are four orders of magnitude in $U$ (see Equation~\eqref{eq:U}), corresponding to roughly a factor of five in dust temperature. It is assumed that the dust size distributions $dn/da$ and alignment function $f_{\rm align}\left(a\right)$ remain constant as $U$ is varied. While there is little frequency dependence in $p$ for $\lambda \gtrsim 200\,\mu$m, the curves for different $U$ values diverge sharply at shorter wavelengths. As $U$ increases, the large, aligned grains responsible for the polarization get warmer, pushing the transition to emission dominated by unaligned, stochastically heated grains to shorter wavelengths. These quantitative predictions can be tested especially in comparison to two-component models, which typically posit that warmer, unaligned carbonaceous grains dominate the dust emission near the peak of the SED \citep[e.g.,]{Draine+Fraisse_2009, Guillet+etal_2018}. Hence, these models predict lower polarization fractions on the Wien side of the dust SED than the present model \citep[cf.,][Figure~15]{Guillet+etal_2018}.

While limited in their ability to access high frequencies, ground-based experiments are poised to tighten constraints on the polarized dust SED at submillimeter and millimeter wavelengths. The Simons Observatory is forecast to improve measurements on the index $\beta$ of the dust opacity law between 93 and 280\,GHz by a factor of $\sim 2$ relative to current constraints \citep{Hensley+etal_2022}. This is sufficient to determine whether the current $\Delta\beta = 0.05\pm0.03$ measured between total and polarized intensity \citep{Planck_2018_XI} is a real difference or a statistical fluctuation. Prime-Cam on the Fred Young Submillimeter Telescope will be capable of polarimetry at 850\,GHz, with discriminating among dust models as one of its key science goals \citep{CCAT_2022}.

Stratospheric and satellite missions are uniquely capable of accessing dust polarization at frequencies where it is the brightest. With an order of magnitude improvement on the Planck sensitivity to dust polarization, the LiteBIRD satellite will furnish detailed measurements of the polarized dust SED to frequencies as high as 448\,GHz following its expected late 2020s launch \citep{LiteBIRD_2022}. The proposed BLAST Observatory \citep{Lowe_2020} and the Probe of Inflation and Cosmic Origins (PICO) Probe concept \citep{Hanany_2019} would both yield measurements of the polarized dust SED at high frequencies (up to 1.7\,THz and 799\,GHz, respectively) and emphasize tests of the one component hypothesis as key science. 

Unfortunately, none of these future or proposed measurements extend to the Wien side of the dust SED. While SOFIA/HAWC+ has provided information on dust in bright, dense regions at these wavelengths \citep[e.g.,][]{Santos_2019, Tram_2021}, it lacks the sensitivity to measure diffuse media. FIR polarimetry on a cryogenic space telescope could access diffuse lines of sight at these wavelengths, providing a definitive test of whether emission near the peak of the dust SED is produced by aligned or unaligned grains and thus of the one versus two component hypothesis.

One-component models require all polarization features to arise from a single grain population and so make specific predictions about the relative strengths of spectroscopic features and continuum polarization. \citet{Draine+Hensley_2021c} found that the adopted properties of astrodust predict a ratio of 10\,$\mu$m to V-band polarization of $2.1\pm0.3$\%, though lower values are possible if the astrodust model overpredicts the true MIR total extinction (see discussion at the end of Section~\ref{subsec:extpol}). Likewise, \citet{Draine+Hensley_2021a} found that the astrodust model predicts polarization in the 3.4\,$\mu$m feature to be 0.016 of that at 10\,$\mu$m, just below current observational limits \citep{Chiar+Tielens_2006}, provided that the 3.4\,$\mu$m absorption is limited to material within a distance $0.010\,\mu$m of the grain surface. Thus new spectropolarimetric measurements of the 3.4 and 9.7\,$\mu$m features are a key test of both the astrodust model as well as the one component hypothesis in general.

One-component models also require that optical polarized extinction and FIR polarized emission arise from the same grain population. The presence of only one grain type producing polarization allows measurements of the amount of optical polarization per unit FIR polarized emission to be translated into constraints on dust properties like shape and porosity with minimal assumptions \citep{Draine+Hensley_2021c}. However, current constraints on this ratio of optical to FIR polarization from \citet{Panopoulou+etal_2019}, based on a single region of the sky with a particularly high polarization fraction, and \citet{Planck_2018_XII}, based on a full-sky analysis, are not fully consistent, suggesting potential spatial variation of this ratio or possibly limitations of comparing pencil beam observations toward stars with degree-scale emission at submillimeter wavelengths. Future efforts to clarify the numerical value of this ratio, extend it to other frequencies, and probe its spatial variation will be powered both by new starlight polarization surveys like PASIPHAE \citep{PASIPHAE} and sensitive new ground-based polarimetry with experiments like the Simons Observatory \citep{Hensley+etal_2022}. These analyses promise not only tests of the one component hypothesis but, if it holds, a window into the variation of the physical properties of dust in different regions of the Galaxy.

\subsubsection{The Nature of Interstellar Nanoparticles}

The MIR emission features attest to the presence of aromatic hydrocarbon nanoparticles that can attain the requisite temperatures to emit at MIR wavelengths through transient heating. The large number of confirmed diffuse interstellar bands (DIBs) points to the presence of many large molecules, which, aside from $C_{60}^+$, have eluded identification. Phenomena like extended red emission, blue luminescence, and the AME are all attributed to interstellar nanoparticles, and particles of this size contribute much of the UV extinction (see Figure~\ref{fig:size_analysis}).

Nanoparticle fragments are an expected byproduct of grain-grain collisions and other dust destruction processes, potentially producing a nanoparticle population that varies significantly with interstellar environment. The size distributions derived in Section~\ref{sec:results} suggest that most of the dust surface area is in the nanoparticles (see Table~\ref{table:quantities}), and so gas phase accretion and photoelectric heating will be dominated by these grains. Given these considerations, we view modeling of the interstellar nanoparticle population as an important frontier and identify here a few key directions for future study.

It is plausible that many, perhaps all, of the DIBs arise in specific nanoparticle species, which, in addition to the PAHs, may include particles of nanosilicate, metallic Fe, or Fe oxide materials. At this time the only spectroscopic evidence is provided by the PAH emission features and the identification of DIBs contributed by C$_{60}^+$ \citep{Campbell+etal_2015, Cordiner+etal_2019}. Identification of some of the other DIBs would pin down the abundances of specific nanoparticles. In a similar vein, individual aromatic molecules have now been identified in dense molecular gas via their rotational spectra \citep{McGuire+etal_2021}. It is unclear whether any one PAH molecule is sufficiently abundant in diffuse gas to be detected by similar means \citep[e.g.,][]{Ali-Haimoud_2014, Ali-Haimoud+etal_2015}, but discovery of any so-called grandPAHs \citep{Andrews+etal_2015} would be an important step forward.

Particles of nanosilicate, if present, may be detectable through emission in the 9.7\,$\mu$m and 18\,$\mu$m features following single photon heating events \citep{Li+Draine_2001}. Metallic Fe particles larger than $\simeq$4.5\,\AA\ in radius are stable to sublimation in the interstellar radiation field \citep{Hensley+Draine_2017a} and, if present, should contribute opacity at microwave wavelengths due to magnetic dipole absorption \citep{Draine+Hensley_2013}. MIR spectra from JWST will likely reveal new features and pose stringent demands on modeling of the continuum emission, thereby constraining models of emission from nanoparticles of various compositions.

The smallest nanoparticles will contribute spinning dust emission unless the electric dipole moment perpendicular to the spin axis vanishes. Thus, the observed ``spinning dust'' spectrum constrains the population of nanoparticles. Observing variations in the frequency spectrum of the AME and understanding the drivers of theses changes will shed light on the nature of the carriers of the emission, the physics of their rotational excitation, and the properties of the interstellar environments in which they reside \citep{Casassus+etal_2021, CepedaArroita+etal_2021}.

If interstellar nanoparticles are aspherical and systematically aligned with the local magnetic field, then emission and absorption from them will be polarized. We have argued that the quantization of energy levels in nanoparticles greatly suppresses the conversion of rotational kinetic energy to heat and thus the alignment process, and we predict negligible polarization from grains of size $a \lesssim 20$\,\AA\ \citep{Draine+Hensley_2016}. This is consistent with strong upper limits on polarization in the DIB features \citep{Cox+etal_2011} and the AME \citep{Macellari+etal_2011, Planck_2015_XXV, GenovaSantos+etal_2017, Herman_2022}. There has however been a reported detection of polarization in the 11.3\,$\mu$m PAH emission feature \citep{Zhang+etal_2017}. Corroboration with other PAH emission features, and continued searches for polarization associated with interstellar nanoparticles in general, is of great importance for its potential to challenge existing theory and to provide new ways of constraining the properties of these grains.

\subsubsection{Towards Statistical Inference of Dust Properties}
\label{subsubsec:stats}

The principal goal of this work is putting forward a model of interstellar dust based on physical material properties that reproduces the phenomenology of dust in the diffuse ISM. We do not argue that this is the unique model necessitated by the observations, nor do we attempt to make statistical statements about the material properties of interstellar grains based on our fits. The optimization described in Section~\ref{sec:data_model} allows us to derive the best fit size distributions and alignment functions of grains assuming fixed material properties. However, as we emphasize throughout Section~\ref{sec:results}, imperfections in the adopted material properties can lead to artifacts in the size distributions and alignment functions. To make robust statistical inferences on the properties of interstellar grains, we would need to simultaneously fit the dielectric functions themselves.

Unfortunately, such an approach is not feasible with our current tools for computing cross sections and temperature distributions. These tools would first need to be optimized for parallel computing or else substantially accelerated given long wall clock times. They also require additional numerical robustness to avoid failures in the limits of extreme dielectric functions or large grains. Currently, such issues are often addressed with manual intervention to identify troublesome regimes and tweak default parameters. If these issues can be overcome, it is possible, and indeed the objective of future work, to infer dust dielectric functions directly from observations in a global Bayesian inference scheme.

\subsubsection{Variation of Dust Properties}
This work is concerned with reproducing the properties of dust in the typical diffuse ISM of the Milky Way. However, dust properties are known to vary with environment. A wide range of extinction laws is observed within the Galaxy \citep[e.g.,]{Fitzpatrick+Massa_1986, Cardelli+Clayton+Mathis_1988, Udalski_2003, Nataf+etal_2013, Schlafly+etal_2017, Fitzpatrick+etal_2019}, as well as within the LMC and SMC \citep[e.g.,][]{Misselt+Clayton+Gordon_1999, Gordon+etal_2003, Hagen_2017}. Variations in the wavelength dependence of optical extinction observed in the Galaxy have an unexpected correlation with the slope of the FIR emission \citep{Schlafly+etal_2016, Zelko+Finkbeiner_2020}. The SMC exhibits unusually large submillimeter emission not easily accounted for with Galactic dust models \citep{Bot_2010, Israel_2010, Planck_Early_XVII, Draine+Hensley_2012}. The wavelength of maximum polarized extinction varies from sightline to sightline in the Galaxy \citep{Serkowski+etal_1975, Wilking+etal_1982, Clayton+Mathis_1988}. The 2175\,\AA\ feature profile has apparent variations in its width but not central wavelength \citep{Fitzpatrick+Massa_1986}. The MIR emission features are observed to have a range of relative strengths depending on environment \citep{Smith+etal_2007, Galliano_2008, Lai_2020}. In dense cloud environments, grains acquire ice mantles \citep{Boogert_2015} and may grow to large sizes \citep{Pagani_2010, Steinacker_2015}. A single, static model of interstellar dust does not suffice.

Some of these variations can likely be understood through changes in the grain size distribution \citep{Weingartner+Draine_2001}. Others, however, may require incorporation of other materials into the model or require a description of how the optical properties of ``astrodust'' change with different levels of gas phase depletion, metallicity, or other environmental factors. While beyond the scope of the present study, answering these questions in the context of physical models---like the one presented here or the THEMIS model---is an important step towards a quantitative understanding the evolution of dust and metals in the ISM.

\section{Conclusions}
\label{sec:summary}
The principal conclusions of this work are as follows:

\begin{enumerate}
    \item We have presented a model of interstellar dust based on a dominant single grain material---``astrodust''---plus a population of PAHs that simultaneously reproduces the observed extinction and emission from dust, both total and polarized, while respecting abundance constraints. Large ($a \gtrsim 0.1\,\mu$m) carbonaceous grains are not required to reproduce any observed properties of dust in the diffuse ISM.
    \item The success of the astrodust model in reproducing observations of dust in the diffuse ISM is consistent with destruction/homogenization of injected stardust on timescales short compared to the $\sim$2.5\,Gyr baryon turnover time, with depletion of selected elements from the gas phase to build new grain material. While pristine stardust grains in the ISM may be predominantly silicate or carbonaceous in nature, we argue that the vast majority of large grains are well-described by a single composition.
    \item We make publicly available\footnote{\url{https://dataverse.harvard.edu/dataverse/astrodust}} all data underlying the model, including tables of opacities and routines for computing extinction, scattering, emission, and polarization from the model as a function of dust column and intensity of the illuminating radiation.
\end{enumerate}

While we have identified a number of refinements that can be made to the specific model presented here, we conclude that an idealized model consisting of PAH nanoparticles plus a dominant population of grains with a uniform ``astrodust'' composition is consistent with current constraints on dust in the ISM. Future observations, particularly polarimetry, will be key for testing this hypothesis and further elucidating the nature of interstellar grains.

\section*{Acknowledgments}
We thank the anonymous referee and Simone Bianchi for helpful comments that improved this manuscript. BSH acknowledges support from the NASA TCAN grant No. NNH17ZDA001N-TCAN and, during the earliest stages of this study, a National Science Foundation Graduate Research Fellowship under Grant No. DGE-0646086. This research was supported in part by NSF Grant AST-1908123. This research was carried out in part at the Jet Propulsion Laboratory, California Institute of Technology, under a contract with the National Aeronautics and Space Administration.

\software{Astropy \citep{astropy:2013, astropy:2018}, DustEM \citep{Compiegne+etal_2011}, Matplotlib \citep{Matplotlib}, NumPy \citep{NumPy}, SciPy \citep{SciPy}, SpDust \citep{Ali-Haimoud+Hirata+Dickinson_2009, Silsbee+AliHaimoud+Hirata_2011}}

\bibliography{refs}
\end{document}